\documentclass[aps,prapplied,reprint,longbibliography,groupedaddress]{revtex4-1}

\usepackage{color,amsmath,amssymb,graphicx}
\usepackage{threeparttable}

\begin{document}


\title{Unveiling the properties of metagratings via a detailed analytical model for synthesis and analysis}


\author{Ariel Epstein}
\email[]{epsteina@ee.technion.ac.il}
\affiliation{Andrew and Erna Viterbi Faculty of Electrical Engineering, Technion - Israel Institute of Technology, Haifa 3200003, Israel}

\author{Oshri Rabinovich}
\affiliation{Andrew and Erna Viterbi Faculty of Electrical Engineering, Technion - Israel Institute of Technology, Haifa 3200003, Israel}


\date{\today}

\begin{abstract}
We present detailed analytical modelling and in-depth investigation of wide-angle reflect-mode metagrating beam splitters. These recently introduced ultrathin devices are capable of implementing intricate diffraction engineering functionalities with only a single meta-atom per macro-period, making them considerably simpler to synthesize than conventional metasurfaces. 
We extend upon recent work and focus on electrically-polarizable metagratings, comprised of loaded conducting wires in front of a perfect elecric conductor, excited by transverse-electric polarized fields, which are more practical for planar fabrication. The derivation further relates the metagrating performance parameters to the individual meta-atom load, facilitating an efficient semianalytical synthesis scheme to determine the required conductor geometry for achieving optimal beam splitting. Subsequently, we utilize the model to analyze the effects of realistic conductor losses, reactance deviations, and frequency shifts on the device performance, and reveal that metagratings feature preferable working points, in which the sensitivity to these non-idealities is rather low. The analytical relations shed light on the physical origin of this phenomenon, associating it with fundamental interference processes taking place in the device. These results, verified via full-wave simulations of realistic physical structures, yield a set of efficient engineering tools, as well as profound physical intuition, for devising future metagrating devices, with immense potential for microwave, terahertz, and optical beam-manipulation applications.
 
\end{abstract}

\pacs{}

\maketitle

\section{Introduction}
\label{sec:introduction}
Metasurfaces have demonstrated in the past few years an exceptional ability to implement a myriad of electromagnetic functionalities, forming highly-efficient ultrathin devices for engineered beam refraction \cite{Pfeiffer2013,Monticone2013,Selvanayagam2013,Epstein2016_3}, reflection \cite{Cui2014,Asadchy2015,Asadchy2016,Estakhri2016_1,Epstein2016_4}, focusing \cite{Lin2014,Aieta2015}, polarization manipulation \cite{Zhao2012,Pfeiffer2014_1,Pfeiffer2014_3,Achouri2015_1,Yin2015}, controlled absorption \cite{Wakatsuchi2013,Asadchy2015_1,Radi2015}, cloaking \cite{Monti2012, Sounas2015, Vellucci2017}, and advanced radiation pattern molding \cite{Pfeiffer2015_1,Epstein2016,Epstein2017,Raeker2016,Raeker2017,Minatti2016_1}, to name a few. These devices are typically designed by prescribing suitable continuous metasurface constituents (\emph{macroscopic} design), implementing a desirable field transformation via the corresponding generalized sheet transition conditions (GSTCs) \cite{Kuester2003, Tretyakov2003, Epstein2016_2,Estakhri2016}. Subsequently, the continuous design specifications are discretized into subwavelength unit cell sizes, and realized using appropriate polarizable particles (\emph{microscopic} design).

While numerous efficient semianalytical \emph{macroscopic} design methods were developed in recent years (e.g., \cite{Epstein2014, Pfeiffer2014_3, Epstein2016_3, Ranjbar2017, Asadchy2016,Estakhri2016_1}), allowing conceptual implementation of advanced field transformations via metasurfaces, translating the latter into physical structures remains a significant challenge. Most of the \emph{microscopic} design schemes rely on full-wave numerical simulations to associate a given subwavelength structure with its equivalent meta-atom constituents, yielding a lookup table that is utilized for general metasurface realization. However, whether in microwave or optical frequencies, bianisotropic metasurfaces, typically necessary for complex beam manipulation, require simultaneous tuning of multiple degrees of freedom at the meta-atom level \cite{Zhao2012, Pfeiffer2014_3, Achouri2015_1, Epstein2016_3, Alaee2015, Alaee2015_1, Odit2016, Kim2016, Asadchy2016_2}; relying on full-wave optimization to engineer each and every meta-atom quickly becomes unreasonable, especially for generally-inhomogeneous metasurfaces (e.g., \cite{Asadchy2016,Epstein2016_4,Epstein2017}).

Very recently, several authors have revisited the problem of perfect reflection, aiming at fully-coupling a plane wave incoming from a given angle to a reflected plane wave propagating towards a desirable (non-specular) direction, based on diffraction grating principles \cite{Sounas2016, Wong2017, Memarian2017, PaniaguaDominguez2017, Radi2017, Wong2017_1}. This problem, which was recently shown to be quite challenging to solve using metasurfaces \cite{Asadchy2016,Estakhri2016_1,Epstein2016_4,DiazRubio2017,Asadchy2017}, turned out to be fully solvable with periodic structures, having only a single or a few subwavelength meta-atoms in each macro-period (whose dimensions are comparable to the wavelength). In contrast to metasurfaces that implement the same functionality, which are comprised of numerous different meta-atoms in a macro-period, these so-called metagratings only require the design of a \emph{single} polarizable particle to achieve an optimal $100\%$ conversion from incident to reflected waves; thus, they substantially overcome the aforementioned microscopic design challenge associated with metasurfaces. 

This complexity reduction is facilitated by the fact that metagratings aim at cancelling a finite number of spurious \emph{propagating} diffraction modes, whereas the metasurfaces implement a prescribed field transformation, which does not allow \emph{any} undesirable diffraction mode (neither propagating nor evanescent) to be excited \cite{Epstein2014_2}. Although this destructive interference mechanism by which efficient diffraction engineering can be achieved is known for many years from the field of dielectric gratings (e.g., \cite{Perry1995,Destouches2005,Ito2013}), a rigorous scheme to determine the optimal grating geometry was absent, and designs were mainly based on physical intuition and numerical optimization. 

In a recent paper, Ra'di \textit{et al.} \cite{Radi2017} developed a rigorous analytical methodology to design metagratings for perfect engineered reflection, based on a periodic array of identical subwavelength particles situated in free space, backed by a perfect electric conductor (PEC). Formulating the fields as a superposition of the fields scattered in the absence of the particle array and the fields generated by the array itself, they found conditions on the required array-PEC separation distance and the effective grid impedance that will guarantee that (1) the specular reflection will destructively interfere with the corresponding Floquet-Bloch (FB) harmonics radiated by the particle array; and (2) all of the incident power will be coupled to a different (prescribed) FB mode. This facilitated perfect reflection via a single-element periodic structure; once the distance between the particle grid and the PEC was determined for given angles of incidence and reflection, the physical structure of the meta-atom was achieved via a simple parametric sweep. Furthermore, it was demonstrated therein that using meta-atoms with more degrees of freedom (e.g., bianisotropic), extends the applicability of such metagratings to additional scenarios.

 \begin{figure*}[htb]
 \includegraphics[width=16cm]{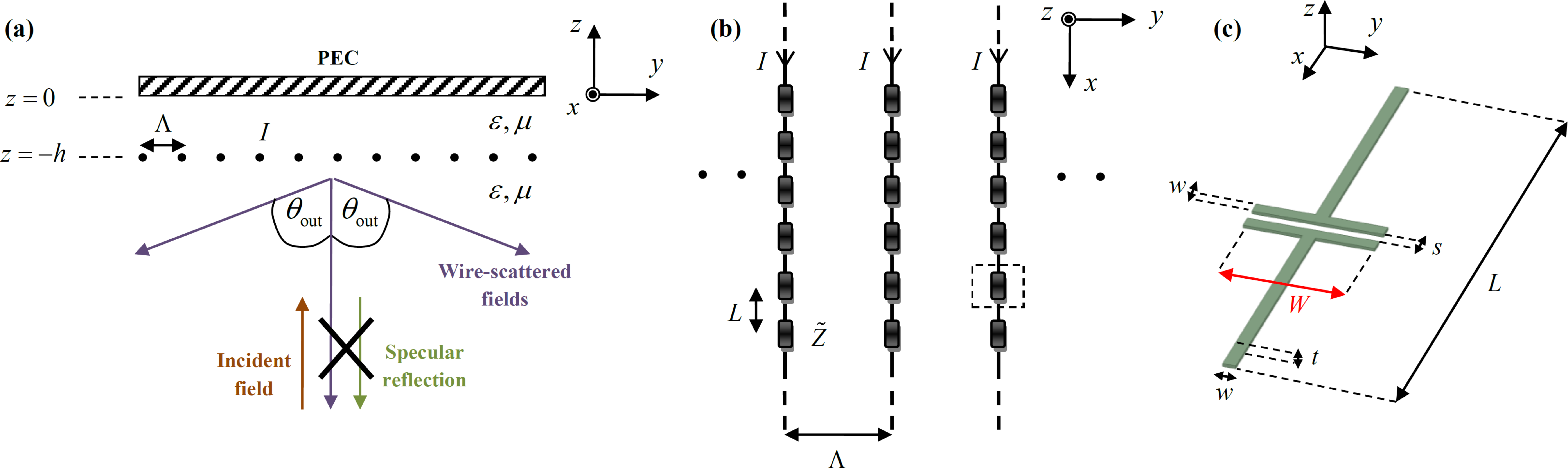}%
 \caption{Physical configuration of the PEC-backed electrically-polarizable beam-splitting metagratings. (a) Side view; $\Lambda$-periodic metagrating separated by $h$ from the PEC, designed to eliminate specular reflection. (b) Top view; distributed impedance per-unit-length $\tilde{Z}$ is formed by finite loads repeating every $L$ along the $x$ axis. (c) Trimetric view of a single electrically-polarizable loaded element [marked by a dashed rectangle in (b)]. trace width, separation, and thickness are given by $w$, $s$, and $t$, respectively; the load impedance is controlled by the capacitor width $W$ (denoted in red).}
 \label{fig:physical_configuration}
 \end{figure*}
 
Recognizing the potential of these novel devices for advanced beam manipulation, we present in this paper a thorough investigation of their fundamental properties. In contrast to \cite{Radi2017}, which utilized magnetically-polarizable particles excited by transverse magnetic (TM) fields, we treat herein electrically-polarizable metagratings, excited by transverse electric (TE) fields (Fig. \ref{fig:physical_configuration}). Focusing on electrically-polarizable particles in the form of loaded conductive wires has two merits. First, such structures are more practical from a realization point of view, as they can be naturally integrated into planar devices, as was vastly demonstrated for microwave, terahertz, and optical metasurfaces (e.g., \cite{Zhao2012, Pfeiffer2014, Kuznetsov2015, Chang2017, Achouri2015_1, Epstein2016}). Second, it allows harnessing of well-established analytical models \cite{Wait1954,Tretyakov2003,Liberal2012} for formulation of efficient and insightful synthesis and analysis schemes.

Indeed, we utilize these models to derive a detailed semianalytical design methodology for reflective metagratings; for simplicity, we focus on perfect wide-angle beam-splitting [Fig. \ref{fig:physical_configuration}(a)], a functionality that was found to be challenging for metasurfaces \cite{Estakhri2016,Estakhri2016_1}, and was mentioned in passing in \cite{Radi2017}. Our derivation goes one step beyond \cite{Radi2017}, deriving analytical expressions for the required individual-wire load impedances. For the capacitive loads suitable for the beam-splitting functionality, we show that this detailed formulation enables analytical determination of the physical dimensions of the required printed-capacitor copper traces, requiring only a single numerical simulation at the frequency of operation.

In addition, we use the detailed analytical model to examine the metagrating performance as a function of load impedance and operating frequency; the model can readily accommodate realistic copper traces with finite conductivity, allowing us to shed light on the role of losses. Our analysis reveals that the metagrating features preferable working points, where the sensitivity to load reactance deviations is low, losses are less pronounced, and the bandwidth is relatively large. These operating conditions are directly linked to fundamental interference processes taking place in the device, as pointed out by the analytical formulation.

These results yield physical insight as well as efficient and intuitive engineering tools for synthesis and analysis of future metagratings, laying the groundwork for practical realization of these devices, and extension of their range of applications.

\section{Theory}
\label{sec:theory}
\subsection{Formulation}
\label{subsec:formulation}
We consider a 2D configuration ($\partial/\partial x=0$) excited by TE-polarized fields ($E_z=E_y=H_x=0$), in which a $\Lambda$-periodic array of loaded conducting wires is situated at $z=-h$ below a PEC, occupying the plane $z=0$ [Fig. \ref{fig:physical_configuration}(a)]. The half-plane $z<0$ is filled with a (passive lossless) homogeneous medium with permittivity $\epsilon$ and permeability $\mu$, defining the wavenumber $k=\omega\sqrt{\mu\epsilon}$ and the wave impedance $\eta=\sqrt{\mu/\epsilon}$ for time-harmonic fields $e^{j\omega t}$. The wires are of width $w\ll\lambda,\Lambda$ and thickness $t\ll w$, where $\lambda=2\pi/k$ is the wavelength at the operating frequency $f=\omega/\left(2\pi\right)$, and are assumed to be uniformly loaded by a distributed impedance per-unit-length of $\tilde{Z}$ [Fig. \ref{fig:physical_configuration}(b)-(c)]. In practice, this distributed impedance is implemented by lumped loads, repeating in a periodic fashion along the $x$-axis with a deep-subwavelength period $L$.

As denoted, our goal is to find the array-PEC distance $h$ and the load impedance $\tilde{Z}$ that yield full and equal coupling of a normally-incident plane wave into two plane waves, reflected towards $\pm\theta_\mathrm{out}$. We start by formulating the total fields in the problem, which can be written as a superposition of the fields in the absence of the wire array, and the fields generated due to the (yet to be determined) current $I$ induced on the wires due to these "external" fields. Each of these sets of fields should comply with the boundary conditions at the PEC, namely, $\left.E_x\left(y,z\right)\right|_{z\rightarrow0^-}=0$. Consequently, the external fields are composed of a normally-incident and normally-reflected plane waves
\begin{equation}
E_x^{\mathrm{ext}}\left(y,z\right) = E_\mathrm{in}\left(e^{-jkz}-e^{jkz}\right),
\label{equ:external_fields}
\end{equation}
where $E_\mathrm{in}$ is the given excitation amplitude. The fields produced by the metagrating are a sum of an infinite array of electric line sources at positions $\left(y,z\right)=\left(n\Lambda,-h\right)$, $n\in\mathbb{Z}$, and their image sources, symmetrically positioned at $\left(y,z\right)=\left(n\Lambda,h\right)$, carrying the same currents with a $\pi$ phase difference. Due to the periodic configuration and the symmetric excitation, the induced currents $I$ are identical for all the wires \cite{Tretyakov2003}, and the corresponding fields are given by
\begin{equation}
\begin{array}{l}
E_x^{\mathrm{wire}}\left(y,z\right) = \\
\,\,\, -\dfrac{k\eta}{4}I\!\!\!\displaystyle\sum\limits_{n=-\infty}^{\infty}\!\!\left\{\!\!
\begin{array}{l}
H_0^{(2)}
	\left[k\sqrt{\left(y-n\Lambda\right)^2+\left(z+h\right)^2}\right] \\
	-H_0^{(2)}
	\left[k\sqrt{\left(y-n\Lambda\right)^2+\left(z-h\right)^2}\right]
	\end{array}
	\!\!\right\},
\end{array}
\label{equ:wire_fields}
\end{equation}
where $H_0^{(2)}\left(\Omega\right)$ is the zeroth-order Hankel function of the second kind.

To evaluate the fields generated by the wires at ${z\neq-h}$, we utilize the Poisson formula \cite{Tretyakov2003}, stating that for a given function $f\left(l\right)$
\begin{equation}
\sum\limits_{n=-\infty}^{\infty}f\left(n\Lambda\right)
=\sum\limits_{m=-\infty}^{\infty}\int\limits_{-\infty}^{\infty}\frac{dl}{\Lambda}f\left(l\right)e^{-j\frac{2\pi m}{\Lambda}l}.
\label{equ:Poisson_formula}
\end{equation}
Using Eq. \eqref{equ:Poisson_formula} with $f\left(l\right)=H_0^{(2)}\left[k\sqrt{\left(y-l\right)^2+\left(z\pm h\right)^2}\right]$, and considering the Fourier transform of the Hankel function is given by \cite[Eqs. (5.4.33)-(5.4.35)]{FelsenMarcuvitz1973}
\begin{equation}
\int\limits_{-\infty}^{\infty}\!\!\!dl H_0^{(2)}\!\!
	\left[\!k\sqrt{\left(y-l\right)^2+\left(z\pm h\right)^2}\right]\!\!e^{-jk_tl}\!=\!2\frac{e^{-jk_ty}e^{-j\beta\left|z\pm h\right|}}{\beta},
\label{equ:Fourier_transform_Hankel}
\end{equation}
where $\beta=\sqrt{k^2-k_t^2}$, $\Im\left\{\beta\right\}\leq0$, Eq. \eqref{equ:wire_fields} can be written as \cite{Tretyakov2003}
\begin{equation}
\begin{array}{l}
\vspace{0.5mm}
E_x^{\mathrm{wire}}\left(y,z\right) = \\
\,\,\, -\dfrac{k\eta}{2\Lambda}I\!\!\!\displaystyle\sum\limits_{m=-\infty}^{\infty}\!\!\!
e^{-j\frac{2\pi m}{\Lambda}y}\frac{e^{-j\beta_m\left|z+h\right|}-e^{j\beta_m\left(z-h\right)}}{\beta_m},
\end{array}
\label{equ:wire_fields_Poisson}
\end{equation}
where $\beta_m=\sqrt{k^2-\left(2\pi m/\Lambda\right)^2}$, $\Im\left\{\beta_m\right\}\leq0$. We can now observe that the interaction of the external fields with the periodic wire array gives rise to a series of scattered FB harmonics, where the $m$th term of the summation in Eq. \eqref{equ:wire_fields_Poisson} corresponds to the $m$th FB mode. 

The total electric fields are thus given by $E_x^\mathrm{tot}\left(y,z\right)=E_x^\mathrm{ext}\left(y,z\right)+E_x^\mathrm{wire}\left(y,z\right)$, and the tangential magnetic fields can be readily derived from them via Maxwell's equations for this TE case, reading ${H_y\left(y,z\right)=-\frac{1}{jk\eta}\frac{\partial}{\partial z}E_x\left(y,z\right)}$.

In the framework of our detailed analysis, we strive to tie the physical structure of the meta-atom (loaded wire) to the design requirements. To this end, we recall that the relation between the total fields at the wire position and the induced currents is given by the distributed impedance $\tilde{Z}$ via Ohm's law, ${\left.E_x^{\mathrm{tot}}\left(y,z\right)\right|_{\left(y,z\right)\rightarrow\left(0,-h\right)}=\tilde{Z}I}$ \cite{Tretyakov2003}. In order to write this expression explicitly, due to the divergence of the Hankel function at $\left(y,z\right)\rightarrow\left(0,-h\right)$, we have to refine our approximation of the current-carrying wire as a line source of infinitesimal radius, and take into account the actual wire dimensions [Fig. \ref{fig:physical_configuration}(c)]. As $t\ll w\ll\lambda$, we can use the flat wire model in \cite{Tretyakov2003}, treating the wire as a conducting cylinder of effective radius $r_\mathrm{eff}=w/4$. Consequently, using Eqs. \eqref{equ:external_fields} and \eqref{equ:wire_fields} we can write Ohm's law as
\begin{equation}
\begin{array}{l}
\vspace{1.5mm}
\tilde{Z}I = 2jE_\mathrm{in}\sin\left(kh\right) \\
\,\,\, -\dfrac{k\eta}{4}I H_0^{(2)}\left(kr_\mathrm{eff}\right)-\dfrac{k\eta}{4}I\!\!\!\!\! \displaystyle\sum\limits_{\scriptsize\begin{array}{c}
							n\!=\!-\infty\\
							n\!\neq\! 0
							\end{array}}^{\infty} \!\!\!\!\!\!H_0^{(2)}\left(k\left|n\Lambda\right|\right) \\
\,\,\, +\dfrac{k\eta}{4}I\!\!\!\displaystyle\sum\limits_{n=-\infty}^{\infty}\!\!\!\!H_0^{(2)}
	\left[k\sqrt{\left(n\Lambda\right)^2+\left(2h\right)^2}\right],
\end{array}
\label{equ:Ohms_law}
\end{equation}
from which the current induced by the applied fields can be evaluated, for a given $\tilde{Z}$. Alternatively, Eq. \eqref{equ:Ohms_law} can be used to assess the required $\tilde{Z}$ to obtain a certain induced current.

Subsequently, we follow \cite{Tretyakov2003} to develop Eq. \eqref{equ:Ohms_law} into a more useful format, expressing the required $\tilde{Z}$ to yield a prescribed $E_\mathrm{in}/I$ ratio (to be derived in Subsections \ref{subsec:specular_reflection} and \ref{subsec:beam_splitting}). In particular, as $w\ll\lambda$, the second term in the right-hand side (RHS) can be approximated by the asymptotic expression of the Hankel function for small arguments \cite[Eq. (9.1.8)]{AbramowitzStegun1970}; the third term can be expanded using \cite[Eq. (8.522)]{GradshteinRyzhik2015}; and for the fourth term, we can apply again the Poisson formula [Eqs. \eqref{equ:Poisson_formula} and \eqref{equ:Fourier_transform_Hankel}]. These transformations lead to
\begin{equation}
\begin{array}{l}
\vspace{0.5mm}
\tilde{Z} = 2j\dfrac{E_\mathrm{in}}{I}\sin\left(kh\right) \\
\vspace{1mm}
\,\,\, -\dfrac{\eta}{2\Lambda}\left(1-e^{-2jkh}\right) +j\dfrac{k\eta}{2\pi}\log\dfrac{2\pi r_\mathrm{eff}}{\Lambda}\\
\,\,\, -k\eta\displaystyle\sum\limits_{m=1}^{\infty}\left(\frac{1-e^{-2j\beta_m h}}{\Lambda\beta_m}-j\frac{1}{2\pi m}\right),
\end{array}
\label{equ:Ohms_law_Poisson}
\end{equation}
in which the infinite summation converges very well.

\subsection{Eliminating specular reflection}
\label{subsec:specular_reflection}
As shown in \cite{Radi2017}, with the available degrees of freedom, namely, $h$ and $\tilde{Z}$, we can only eliminate a single FB mode. Thus, to successfully couple all the incident power to the FB modes propagating towards $\pm\theta_\mathrm{out}$, these have to be the only FB modes (other than the fundamental specular reflection) that are propagating. This requirement imposes two constraints on our design. First, the angles $\pm\theta_\mathrm{out}$ should correspond to the $\pm1$ propagating FB modes; following Eq. \eqref{equ:wire_fields_Poisson} this implies that
\begin{equation}
\frac{2\pi}{\Lambda}=k\sin\theta_\mathrm{out}\,\Rightarrow\,\Lambda=\frac{\lambda}{\sin\theta_\mathrm{out}}.
\label{equ:Lambda_period}
\end{equation}
Second, all the other higher-order FB modes ($\left|m\right|\geq2$) should be evanescent, implying, from Eqs. \eqref{equ:wire_fields_Poisson} and \eqref{equ:Lambda_period}, that
\begin{equation}
2\frac{2\pi}{\Lambda}>k\,\Rightarrow\,\theta_\mathrm{out}>30^\circ.
\label{equ:theta_constraint}
\end{equation}

Let us apply these constraints on the field expressions, and write the total fields $E_x^{\mathrm{tot},<}$ below the metagrating ($z<-h$) using Eqs. \eqref{equ:external_fields} and \eqref{equ:wire_fields_Poisson}. These read
\begin{equation}
\begin{array}{l}
\vspace{1mm}
E_x^{\mathrm{tot},<}\left(y,z\right) = E_\mathrm{in}e^{-jkz}-E_\mathrm{in}e^{jkz} \\
\vspace{1mm}
\,\,\, -j\dfrac{\eta I}{\Lambda}\sin\left(kh\right)e^{jkz} \\
\vspace{1mm}
\,\,\, -j\dfrac{\eta I}{\Lambda}\frac{\sin\left(kh\cos\theta_\mathrm{out}\right)}{\cos\theta_\mathrm{out}} e^{jkz\cos\theta_\mathrm{out}}e^{-jky\sin\theta_\mathrm{out}}  \\
\vspace{1mm}
\,\,\, -j\dfrac{\eta I}{\Lambda}\frac{\sin\left(kh\cos\theta_\mathrm{out}\right)}{\cos\theta_\mathrm{out}} e^{jkz\cos\theta_\mathrm{out}}e^{jky\sin\theta_\mathrm{out}}  \\
\,\,\, -j\dfrac{\eta I}{\Lambda}\!\!\!\displaystyle\sum\limits_{\scriptsize\begin{array}{c}
							m\!=\!-\infty\\
							|m|\!\geq\! 2
							\end{array}}^{\infty}\!\!\!
\frac{k\sinh\left(\alpha_m h\right)}{\alpha_m}e^{\alpha_m z}e^{-j\frac{2\pi m}{\Lambda}y},
\end{array}
\label{equ:total_fields_below}
\end{equation}
where we used $\beta_m\triangleq-j\alpha_m$ ($\alpha_m\geq 0$, $\forall \left|m\right|\geq2$) in the terms corresponding to the evanescent modes according to Eq. \eqref{equ:theta_constraint}.

From Eq. \eqref{equ:total_fields_below} it is quite clear that our only means to eliminate the specular reflection (second term in RHS) is to form destructive interference with the fundamental FB mode of the wire-generated fields (third term in RHS) \cite{Radi2017}. Consequently, we are required to tune the physical configuration of Fig. \ref{fig:physical_configuration}(c) such that
\begin{equation}
\dfrac{E_\mathrm{in}}{I}=-j\dfrac{\eta}{\Lambda}\sin\left(kh\right).
\label{equ:specular_elimination_condition}
\end{equation}

\subsection{Perfect beam splitting}
\label{subsec:beam_splitting}
Once we have eliminated specular reflections via Eq. \eqref{equ:specular_elimination_condition}, we should guarantee that all of the incident power indeed couples to the two plane waves propagating towards $\pm\theta_\mathrm{out}$ (i.e., the $\pm1$ FB modes). Although these are the only propagating modes that are left [Eq. \eqref{equ:total_fields_below}], the incident power could be partially absorbed by the metagrating, reducing the device performance; in this subsection, we derive the condition to avoid this undesirable absorption.

In order to ensure that all the incident power is coupled to the two reflected beams, we merely need to require that the net real power crossing a certain plane $z=z_p<-h$ vanishes; this means that the real power incident upon the metagrating is reflected in its entirely. As the $\pm1$ FB modes are the only propagating modes that remain after the elimination of specular reflection, this implies that all the incident power is coupled to these modes; due to the problem symmetry, the same amount of power is coupled to each of these plane waves.

The overall real power crossing the plane $z=z_p<-h$ in one period is defined as
\begin{equation}
P_z^{\mathrm{tot}}\left(z\right)=\frac{1}{2}\!\!\!\int\limits_{-\Lambda/2}^{\Lambda/2} \!\!\! dy \, \Re\left\{E_x\left(y,z\right)H_y^*\left(y,z\right)\right\}.
\label{equ:real_power_definition}
\end{equation}
Due to he problem periodicity, it is sufficient to show that the real power integrated over a single period indeed vanishes to guarantee full coupling as discussed above. Subsequently, the perfect beam-splitting condition $P_z^{\mathrm{tot}}\left(z_p\right)=0$ can be written explicitly by substituting Eq. \eqref{equ:total_fields_below} (and its $z$-derivative, corresponding to the tangential magnetic fields) into Eq. \eqref{equ:real_power_definition}, integrating, and equating to zero. This yields a second condition on the metagrating parameters, namely,
\begin{equation}
\begin{array}{l}
\vspace{1mm}
\Im\left\{\dfrac{E_{\mathrm{in}}}{I}\right\}\sin\left(kh\right) + \dfrac{\eta}{2\Lambda}\sin^2\left(kh\right) = \\
\,\,\,\,\,\,\,\,\,\, -\dfrac{\eta}{\Lambda\cos\theta_\mathrm{out}}\sin^2\left(kh\cos\theta_\mathrm{out}\right)
\end{array}
\label{equ:full_coupling_condition}
\end{equation}
Note that as we consider a passive lossless medium $\{\epsilon,\mu\}\in\mathbb{R}$, the perfect beam-splitting condition is independent of the choice of $z_p$.

Substituting the specular reflection elimination condition Eq. \eqref{equ:specular_elimination_condition} into Eq. \eqref{equ:full_coupling_condition}, still considering a passive lossless medium $\{k,\eta\}\in\mathbb{R}$, yields 
\begin{equation}
\mathcal{E}=\cos\theta_\mathrm{out}\sin^2\left(kh\right)-2\sin^2\left(kh\cos\theta_\mathrm{out}\right)=0,
\label{equ:h_condition}
\end{equation}
which is a nonlinear equation from which the required wire-PEC separation distance $h$ can be numerically/graphically evaluated, setting our first degree of freedom. Compared with the analogous Eq. (4) of \cite{Radi2017}, we can observe that the interference terms (trigonometric functions with arguments $kh$ and $kh\cos\theta_\mathrm{out}$) feature now sines instead of cosines (due to difference between image theory for TE and TM polarized sources), and the perfactors correspond to the wave impedances of the various propagating modes (note that herein we have three distinct propagating FB modes).

After fixing $h$ following Eq. \eqref{equ:h_condition}, Eqs. \eqref{equ:specular_elimination_condition} and \eqref{equ:full_coupling_condition} can be substituted into Eq. \eqref{equ:Ohms_law_Poisson} to obtain an explicit expression for the distributed impedance $\tilde{Z}$, reading
\begin{equation}
\begin{array}{l}
\vspace{0.5mm}
\tilde{Z} = -j\dfrac{\eta}{\Lambda}\left[\dfrac{\sin\left(2kh\right)}{2}+\dfrac{\sin\left(2kh\cos\theta_\mathrm{out}\right)}{\cos\theta_\mathrm{out}}\right] \\
\vspace{1mm}
\,\,\,  +j\dfrac{k\eta}{2\pi}\left(1+\log\dfrac{2\pi r_\mathrm{eff}}{\Lambda}\right)\\
\,\,\, -j\dfrac{\eta}{\Lambda}\displaystyle\sum\limits_{m=2}^{\infty}\left[\frac{k\left(1-e^{-2\alpha_m h}\right)}{\alpha_m}-\frac{k\Lambda}{2\pi m}\right],
\end{array}
\label{equ:Z_tilde_condition}
\end{equation}
setting our second degree of freedom.

The benefits of providing direct access to the individual wire load in our synthesis scheme are apparent already from a brief look at Eq. \eqref{equ:Z_tilde_condition}. It can be readily verified that the RHS is purely imaginary; this indicates that in order to have full coupling of the incident plane wave into the two symmetrical diffraction modes, the wire should be loaded by a purely reactive impedance. This is consistent with our previous observation that only losses could prevent perfect beam-splitting once the specular reflection elimination condition of Eq. \eqref{equ:specular_elimination_condition} is satisfied, and thus should ideally be avoided.

\section{Results and Discussion}
\label{sec:results}
\subsection{Synthesis}
\label{subsec:synthesis}
\begin{figure}[tbh]
 \includegraphics[width=7cm]{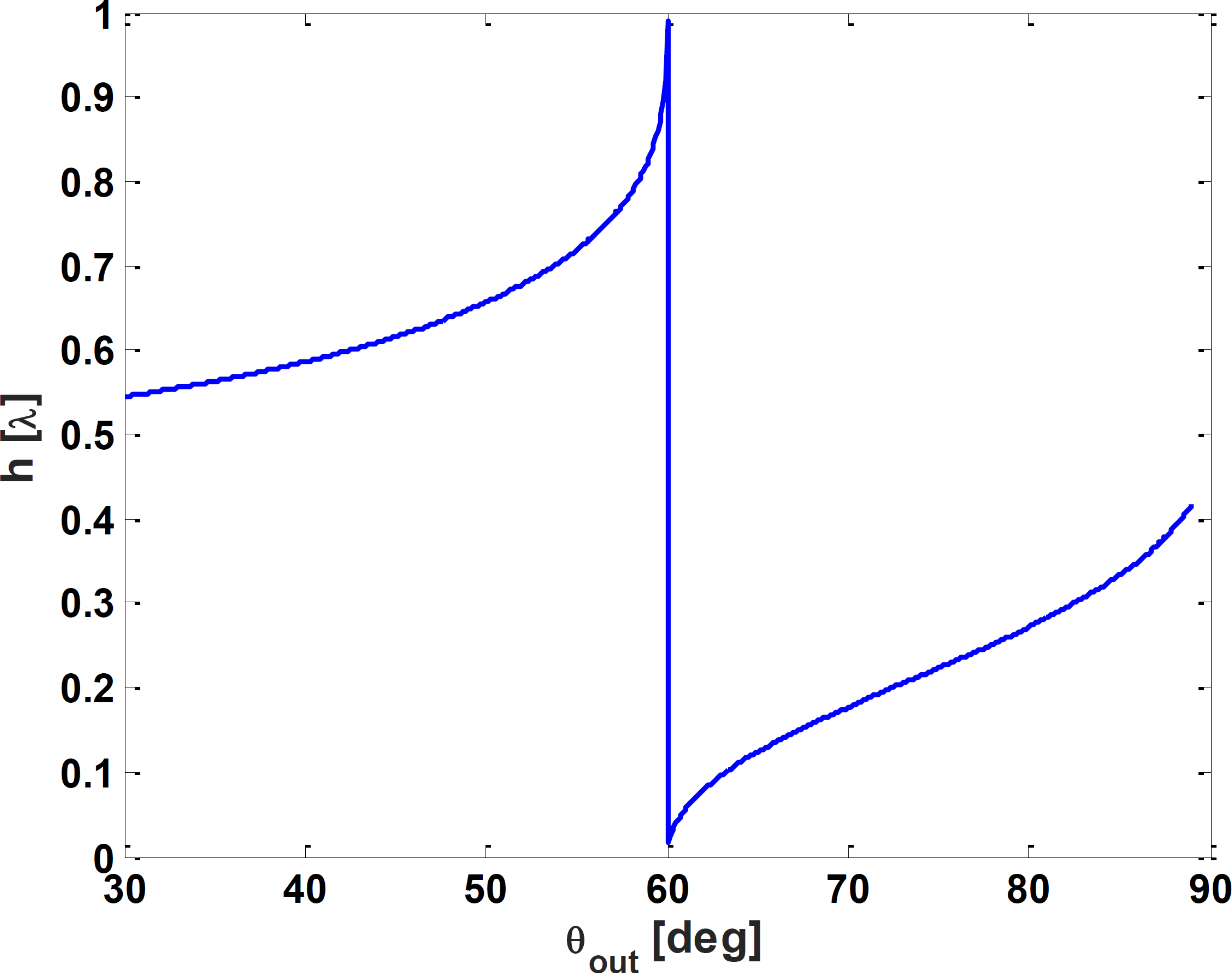}%
 \caption{Required wire-PEC separation as a function of the splitting angle, obtained from Eq. \eqref{equ:h_condition}.}
 \label{fig:hCondition}
 \end{figure}

We first use the developed formalism to demonstrate an efficient way for synthesizing perfect metagrating beam splitters. To this end, for a given desirable $\theta_\mathrm{out}$, we find (via a simple numerical MATLAB code) the separation distance $h$ that minimizes $\mathcal{E}$ of Eq. \eqref{equ:h_condition}. The optimal wire-PEC distance is presented in Fig. \ref{fig:hCondition} as a function of the splitting angle, where we have chosen the smallest $h$ satisfying Eq. \eqref{equ:h_condition} for each $\theta_\mathrm{out}$. This a universal curve, which is valid for all operating frequencies (note that $h$ is expressed in wavelength units). Therefore, we may conclude that it is feasible to implement all the possible beam splitters with metagrating devices whose thickness is less than the operating wavelength.

Subsequently, to evaluate the required distributed impedance (the other degree of freedom we need to set), we substitute these optimal $h$ values (Fig. \ref{fig:hCondition}) into Eq. \eqref{equ:Z_tilde_condition}, considering the suitable metagrating period $\Lambda$ for each splitting angle [Eq. \eqref{equ:Lambda_period}]. 
For a fixed conductor width $w$ [Fig. \ref{fig:physical_configuration}(c)], typically limited by manufacturing constraints, this design curve \emph{does} depend on the operation frequency, due to the expression in the second row of Eq. \eqref{equ:Z_tilde_condition} [recall that $r_\mathrm{eff}=w/4$]. Thus, to proceed with our device synthesis, we need to fix $w$, and consider specific operating frequencies.

Throughout this paper, we will consider the printed capacitor geometry presented in Fig. \ref{fig:physical_configuration}(c) for implementing the distributed load 
(the reasons for choosing a distributed capacitance will become apparent shortly). The trace width and trace separation are fixed to ${w=s=3\mathrm{mil}=76.2\mathrm{\mu m}}$ [Fig. \ref{fig:physical_configuration}(c)], following typical fabrication tolerances \cite{Epstein2016,Chen2017}. This structure repeats itself periodically every $L=\lambda/10$ along the $x$-axis, forming an approximately-homogeneous distributed capacitance. The equivalent impedance per-unit-length $\tilde{Z}$ of this formation can be thus tuned by modifying the capacitor width $W$, which is approximately linearly-proportional to the capacitance \cite{Lee2003}.

 \begin{figure}[tbh]
 \includegraphics[width=8cm]{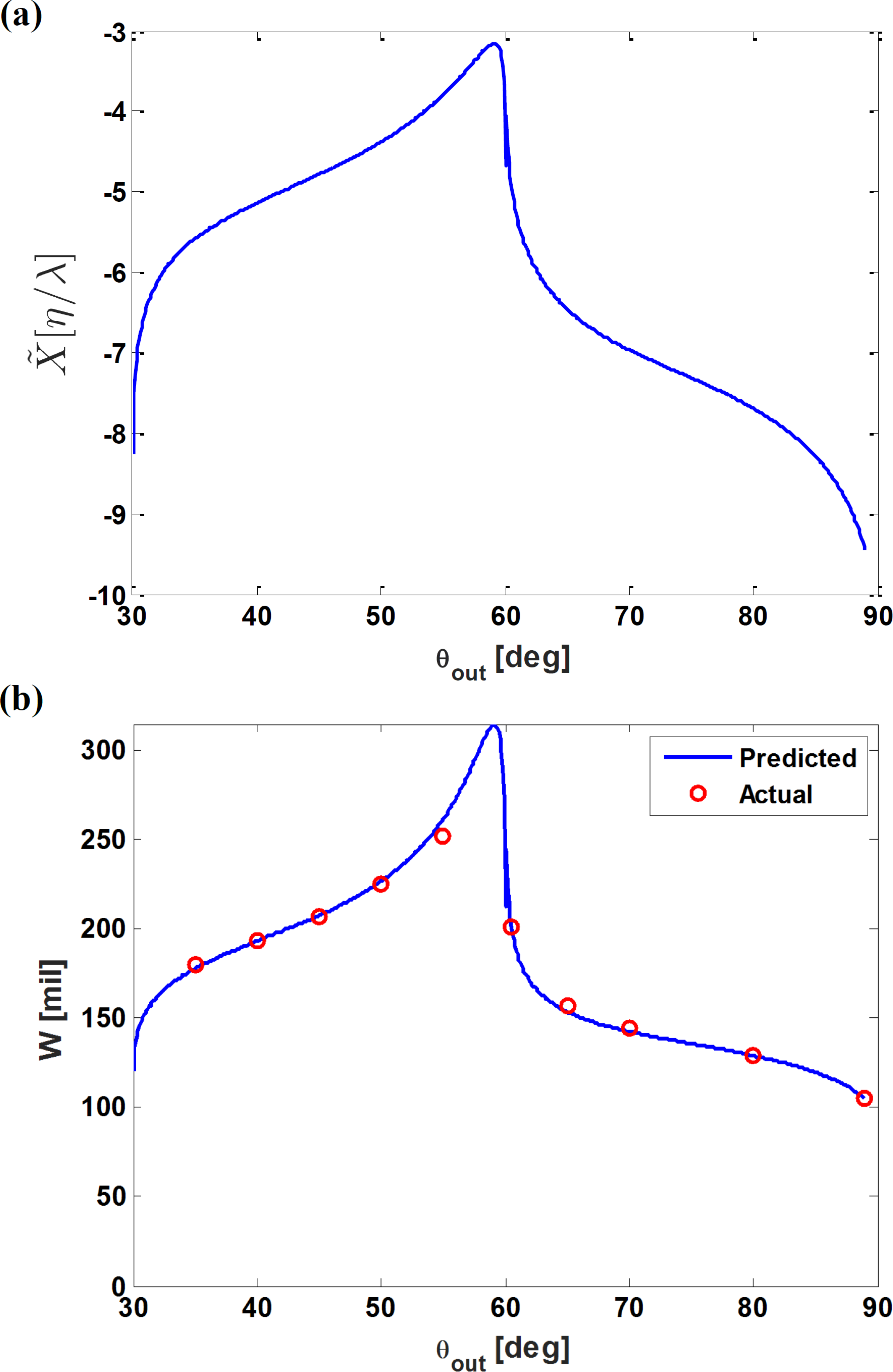}%
 \caption{Load design specifications as a function of the splitting angle, for metagratings operating at $f=10\mathrm{GHz}$. (a) Required distributed reactance $\tilde{X}=\Im\{\tilde{Z}\}$, evaluated from Eq. \eqref{equ:Z_tilde_condition}. (b) Corresponding capacitor width $W$ [Fig. \ref{fig:physical_configuration}(c)], comparing predictions via Eq. \eqref{equ:capacitor_width} (blue solid line) with actual optimal values obtained from full-wave simulations (red circles).}
 \label{fig:load_design_10GHz}
 \end{figure}

Using this geometry, we plot in Fig. \ref{fig:load_design_10GHz}(a) the required distributed reactance $\tilde{X}\triangleq\Im\{\tilde{Z}\}$ as a function of the splitting angle for the operating frequency $f=10\mathrm{GHz}$ ($\lambda\approx30\mathrm{mm}$), obtained from Eq. \eqref{equ:Z_tilde_condition} and the results of Fig. \ref{fig:hCondition}. As can be observed, the required reactance is negative for all considered $\theta_\mathrm{out}$; thus, a capacitive loading is required, given by $C=-1/(2\pi f L \tilde{X})$, which explains the chosen meta-atom geometry [Fig. \ref{fig:physical_configuration}(c)]. 

 \begin{figure*}[htb]
 \includegraphics[width=16cm]{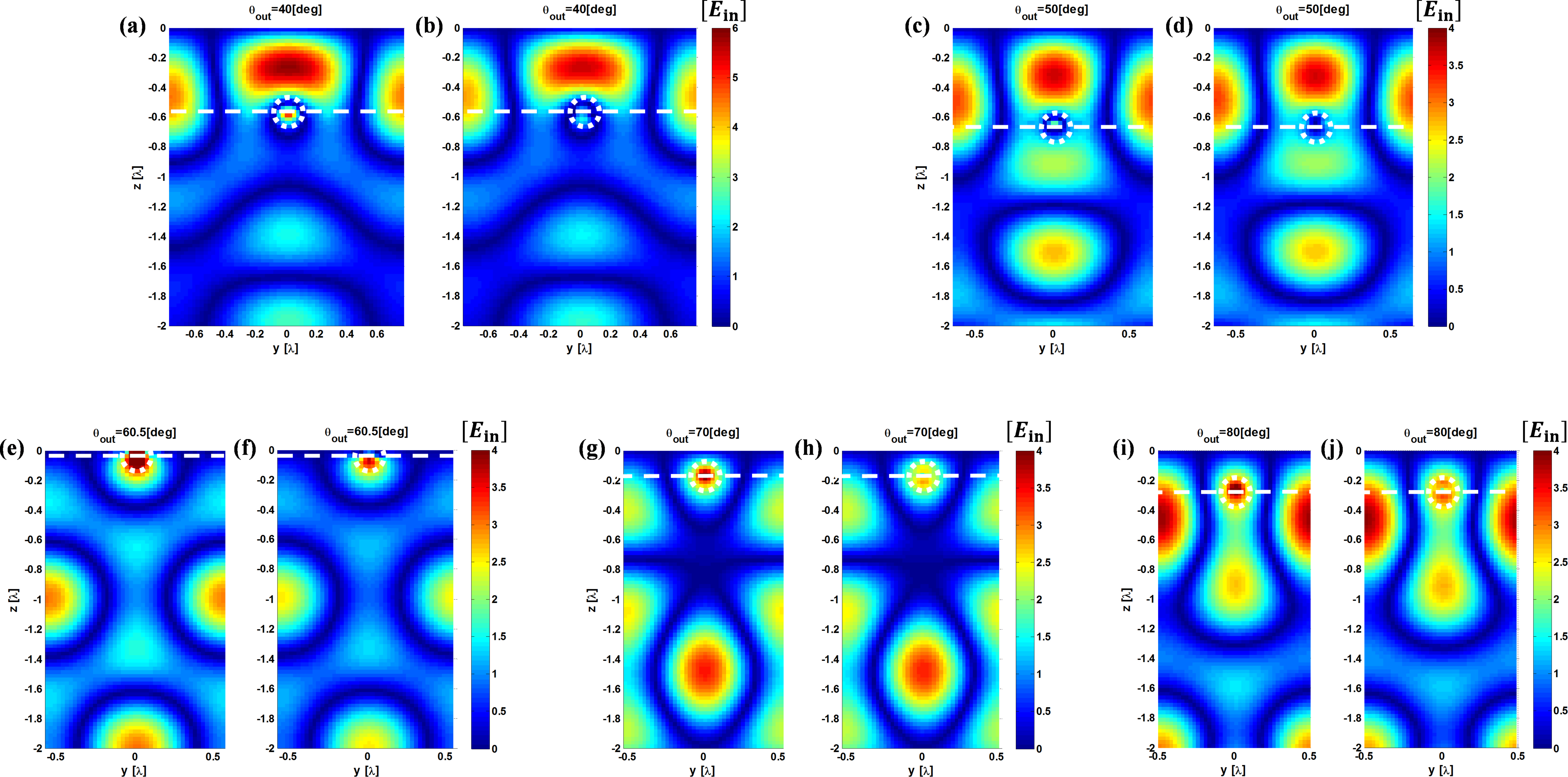}%
 \caption{Electric field distributions $\left|\Re\left\{E_x\left(y,z\right)\right\}\right|$ for beam-splitting metagratings operating at $f=10\mathrm{GHz}$, excited from below with a normally-incident plane wave. Analytical predictions following Eqs. \eqref{equ:external_fields} and \eqref{equ:wire_fields_Poisson} [(a),(c),(e),(g),(i)] are compared to results of full-wave simulations of the realistic loaded wires of Fig. \ref{fig:physical_configuration}(c) with the optimal values of Fig. \ref{fig:load_design_10GHz}(b) [(b),(d),(f),(h),(j)]. A single period $\Lambda=\lambda/\sin\theta_\mathrm{out}$ is shown, for metagratings designed following Eqs. \eqref{equ:h_condition} and \eqref{equ:Z_tilde_condition} for various splitting angles: (a),(b) $\theta_\mathrm{out}=40^\circ$; (c),(d) $\theta_\mathrm{out}=50^\circ$; (e),(f) $\theta_\mathrm{out}=60.5^\circ$; (g),(h) $\theta_\mathrm{out}=70^\circ$; and (i),(j) $\theta_\mathrm{out}=80^\circ$. Dashed horizontal white lines denote the plane $z=-h$ of Eq. \eqref{equ:h_condition}, and a dotted white circle denotes a $0.1\lambda$-diameter region around the metagrating element, within which analytical predictions for uniformly-loaded singular wires are expected to deviate from full-wave simulations of realistic copper traces.}
 \label{fig:fields_10GHz}
 \end{figure*}
 
The last step to obtain a detailed physical realization involves assessing the required capacitor width $W$ that implements the prescribed quasi-static capacitance $C$. To this end, we can use certain analytical approximations for the capacitance of coplanar strips; however, as these do not usually consider residual capacitance formed due to the vertical lines connecting the printed capacitors (i.e., the wire itself), a frequency-dependent correction factor $K_\mathrm{corr}$ should be incorporated into these formulas. Fortunately, as the capacitance is predominantly proportional to the capacitor width $W$, once this correction factor is assessed via full-wave simulations for one working point, it can be used to generate other designs, as long as the operation frequency remains the same. Specifically, we follow \cite[Eq. (7.64)]{Gupta1996}, which for our case of $w=s$ yields the following approximation for the required capacitor width
\begin{equation}
W\approx 2.85K_\mathrm{corr}C \,\left[\dfrac{\mathrm{mil}}{\mathrm{fF}}\right]
\label{equ:capacitor_width}
\end{equation}

We use a commercial finite-elements solver, ANSYS HFSS, to compare the analytical predictions (Section \ref{sec:theory}) with full-wave simulations of the metagrating realization. For a given $\theta_\mathrm{out}$, the simulation domain consists of a PEC at $z=0$ and a loaded-wire meta-atom [Fig. \ref{fig:physical_configuration}(c)] at the corresponding $z=-h$ (Fig. \ref{fig:hCondition}), placed inside a 2D Master-Slave periodic boundary conditions [$\Lambda$-periodic along the $y$-axis and $L$-periodic along the $x$-axis, \textit{cf.} Fig. \ref{fig:physical_configuration}(a),(b)], excited by a Floquet port at $z=-2\lambda$. The standard value of $\sigma=58\times10^6 \mathrm{S/m}$ was used to simulate realistic copper conductivity, further enhancing the fidelity of the simulation results.

First, to evaluate $K_\mathrm{corr}$ at $f=10\mathrm{GHz}$, we consider the configuration corresponding to $\theta_\mathrm{out}=80^\circ$ (chosen arbitrarily), and sweep the capacitor width around the value predicted by Eq. \eqref{equ:capacitor_width} without correction ($K_\mathrm{corr}=1$) to find the actual optimal $W$, which yields the highest power coupling to the $\pm1$ FB modes. The ratio between the uncorrected and the optimal $W$ forms the required correction factor, which is found to be ${K_\mathrm{corr}@10\mathrm{GHz}=0.83}$. 

Next, we use this value with Eq. \eqref{equ:capacitor_width} and the prescribed distributed impedance Fig. \ref{fig:load_design_10GHz}(a) to predict the required capacitor width for all other $\theta_\mathrm{out}$; Figure \ref{fig:load_design_10GHz}(b) presents the required $W$ values (blue solid line) obtained in this manner. Subsequently, for representative split angles in the range $\theta_\mathrm{out}=35^\circ$ to $\theta_\mathrm{out}=89^\circ$, we sweep $W$ in full-wave simulations around the predicted value to find the actual optimal capacitor width; these optima are denoted using red circles in Fig. \ref{fig:load_design_10GHz}(b). As can be observed, excellent agreement between the semianalytical predictions [Eq. \eqref{equ:capacitor_width}] and the optimal values is obtained. This points out another advantage of the detailed analytical model used in this paper, namely, its ability to provide a very good prediction of the optimal physical dimensions of the meta-atom geometry.

\begin{table*}[htb]
\centering
\begin{threeparttable}[b]
\renewcommand{\arraystretch}{1.3}
\caption{Design specifications and simulated performance of beam-splitting metagratings operating at $f=10\mathrm{GHz}$ (corresponding to Figs. \ref{fig:load_design_10GHz} and \ref{fig:fields_10GHz}).}
\label{tab:metagrating_performance_10GHz}
\centering
\begin{tabular}{l|c|c|c|c|c|c|c|c|c|c}
\hline \hline
$\theta_\mathrm{out}$ 
& $35^\circ$ & $40^\circ$ & $45^\circ$ 
& $50^\circ$ & $55^\circ$ & $60.5^\circ$ 
& $65^\circ$ & $70^\circ$ & $80^\circ$ 
& $89^\circ$  \\ 
\hline \hline \\[-1.3em]
	\begin{tabular}{l} $\Lambda [\lambda]$ \end{tabular}
	 & $1.743$ & $1.556$ & $1.414$ 
	& $1.305$ & $1.221$ & $1.149$ 
	& $1.103$ & $1.064$ & $1.016$ 
	& $1.0002$  \\	\hline	
	 \begin{tabular}{l} $h [\lambda]$ \end{tabular}
	 & $0.562$ & $0.586$ & $0.616$ 
	& $0.656$ & $0.718$ & $0.039$ 
	& $0.123$ & $0.176$ & $0.272$ 
	& $0.418$  \\	\hline 
	 \begin{tabular}{l} $W [\mathrm{mil}]$ \end{tabular}
	 & $179.6$ & $193.5$ & $207.0$ 
	& $225.3$ & $252.0$ & $201.8$ 
	& $158.1$ & $144.0$ & $129.0$ 
	& $105.0$  \\	\hline 	  
	 \begin{tabular}{l} Splitting efficiency \end{tabular}
	 & $2\times40.5\%$ & $2\times44.9\%$ & $2\times47.0\%$ 
	& $2\times48.1\%$ & $2\times48.6\%$ & $2\times35.5\%$ 
	& $2\times48.0\%$ & $2\times48.9\%$ & $2\times49.1\%$ 
	& $2\times46.7\%$  \\	\hline 
	\begin{tabular}{l} Specular reflection \end{tabular}
	 & $1.4\%$ & $0.1\%$ & $0.2\%$ 
	& $0.3\%$ & $0.3\%$ & $2.5\%$ 
	& $0.2\%$ & $0.0\%$ & $0.1\%$ 
	& $0.2\%$  \\	\hline 
	 \begin{tabular}{l} Losses \end{tabular}
	 & $17.6\%$ & $10.1\%$ & $5.8\%$ 
	& $3.5\%$ & $2.5\%$ & $26.5\%$ 
	& $3.8\%$ & $2.2\%$ & $1.7\%$ 
	& $6.4\%$  \\		
\hline \hline
\end{tabular}
\end{threeparttable}
\end{table*}

Figure \ref{fig:fields_10GHz} presents the field distributions as obtained from the analytical predictions [Eqs. \eqref{equ:external_fields}, \eqref{equ:wire_fields_Poisson}, \eqref{equ:specular_elimination_condition}, and \eqref{equ:h_condition}] and from full-wave simulations with the realistic metagrating elements of Fig. \ref{fig:physical_configuration}(c) and the optimal capacitor widths of Fig. \ref{fig:load_design_10GHz}(b), for several representative split angles. These plots reflect an excellent agreement between the analytical theory and the simulated actual devices, except for small regions around the meta-atoms (denoted in dotted white circles of diameter $0.1\lambda$), where the uniformly-loaded singular wire model used in the analytical calculations fails to account for the finite-size copper trace geometry used in simulations. 

A closer look reveals that although the predicted and simulated field interference patterns almost-perfectly match, the absolute field amplitudes in the simulated results are lower than the predicted ones (note that the same colorbar scale is used). While for most considered designs these deviations are rather minor, for certain split angles, e.g. for $\theta_\mathrm{out}=60.5^\circ$ [Fig. \ref{fig:fields_10GHz}(e),(f)], the differences are quite significant. This reduction in field amplitude is related to conductor losses, which are taken into account in the simulated realistic design, but were so-far ignored in the analytical model. 

Indeed, as can be observed in Table \ref{tab:metagrating_performance_10GHz}, summarizing the design specifications and simulated performance parameters for metagrating beam-splitters with various split angles (including those presented in Fig. \ref{fig:fields_10GHz}), certain values of $\theta_\mathrm{out}$ are more prone to losses than others. While for most working points a high splitting efficiency is obtained, with more than $2\times45\%$ of the incident power coupled symmetrically to the $\pm1$ FB modes, losses increase when $\theta_\mathrm{out}\rightarrow30^{\circ}$,  $\theta_\mathrm{out}\rightarrow60^{\circ}$, and $\theta_\mathrm{out}\rightarrow90^{\circ}$.
Interestingly, the losses do not increase monotonically with increasing split angle, which implies that the performance reduction in metagratings is not related to impedance mismatch as is the case of Huygens' metasurfaces \cite{Epstein2014, Epstein2014_2, Wong2016, Epstein2016_3, Epstein2016_4, Asadchy2016}, but is rather driven by a different mechanism, yet to be investigated.
Overall, Table \ref{tab:metagrating_performance_10GHz} verifies that the simple single-element periodic metagratings can indeed reach very high splitting efficiencies even for extreme split angles, limited only by losses (note that specular reflection is practically negligible for all cases). 

 \begin{figure}[tbh]
 \includegraphics[width=8cm]{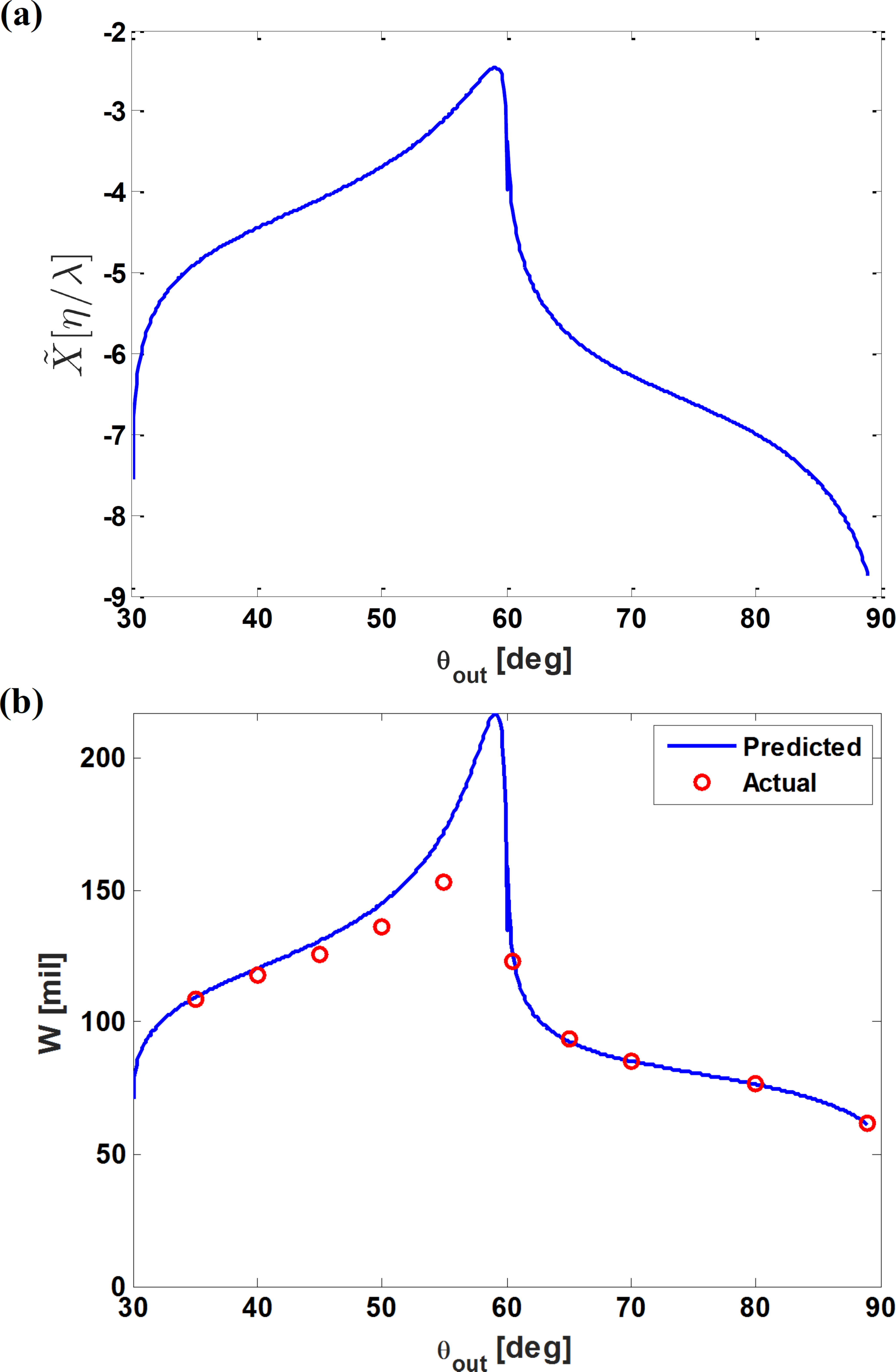}%
 \caption{Load design specifications as a function of the splitting angle, for metagratings operating at $f=20\mathrm{GHz}$. (a) Required distributed reactance $\tilde{X}=\Im\{\tilde{Z}\}$, evaluated from Eq. \eqref{equ:Z_tilde_condition}. (b) Corresponding capacitor width $W$ [Fig. \ref{fig:physical_configuration}(c)], comparing predictions via Eq. \eqref{equ:capacitor_width} (blue solid line) with actual optimal values obtained from full-wave simulations (red circles).}
 \label{fig:load_design_20GHz}
 \end{figure}
 
 \begin{table*}[htb]
\centering
\begin{threeparttable}[b]
\renewcommand{\arraystretch}{1.3}
\caption{Design specifications and simulated performance of beam-splitting metagratings operating at $f=20\mathrm{GHz}$ (corresponding to Fig. \ref{fig:load_design_20GHz}).}
\label{tab:metagrating_performance_20GHz}
\centering
\begin{tabular}{l|c|c|c|c|c|c|c|c|c|c}
\hline \hline
$\theta_\mathrm{out}$ 
& $35^\circ$ & $40^\circ$ & $45^\circ$ 
& $50^\circ$ & $55^\circ$ & $60.5^\circ$ 
& $65^\circ$ & $70^\circ$ & $80^\circ$ 
& $89^\circ$  \\ 
\hline \hline \\[-1.3em]
	\begin{tabular}{l} $\Lambda [\lambda]$ \end{tabular}
	 & $1.743$ & $1.556$ & $1.414$ 
	& $1.305$ & $1.221$ & $1.149$ 
	& $1.103$ & $1.064$ & $1.016$ 
	& $1.0002$  \\	\hline	
	 \begin{tabular}{l} $h [\lambda]$ \end{tabular}
	 & $0.562$ & $0.586$ & $0.616$ 
	& $0.656$ & $0.718$ & $0.039$ 
	& $0.123$ & $0.176$ & $0.272$ 
	& $0.418$  \\	\hline 
	 \begin{tabular}{l} $W [\mathrm{mil}]$ \end{tabular}
	 & $109.1$ & $117.8$ & $126.0$ 
	& $136.5$ & $153.0$ & $123.0$ 
	& $93.8$ & $85.5$ & $76.5$ 
	& $61.5$  \\	\hline 	  
	 \begin{tabular}{l} Splitting efficiency \end{tabular}
	 & $2\times42.3\%$ & $2\times45.8\%$ & $2\times47.6\%$ 
	& $2\times48.6\%$ & $2\times49.0\%$ & $2\times37.8\%$ 
	& $2\times48.5\%$ & $2\times49.1\%$ & $2\times49.3\%$ 
	& $2\times47.4\%$  \\	\hline 
	\begin{tabular}{l} Specular reflection \end{tabular}
	 & $0.8\%$ & $0.2\%$ & $0.1\%$ 
	& $0.0\%$ & $0.0\%$ & $1.7\%$ 
	& $0.0\%$ & $0.0\%$ & $0.1\%$ 
	& $0.2\%$  \\	\hline 
	 \begin{tabular}{l} Losses \end{tabular}
	 & $14.6\%$ & $8.2\%$ & $4.7\%$ 
	& $2.8\%$ & $2.0\%$ & $22.7\%$ 
	& $3.0\%$ & $1.8\%$ & $1.3\%$ 
	& $5.0\%$  \\		
\hline \hline
\end{tabular}
\end{threeparttable}
\end{table*}

To further demonstrate the versatility of our synthesis scheme and analytical model, as well as to verify the observations made after Table \ref{tab:metagrating_performance_10GHz}, we apply the prescribed methodology to design beam splitters at another frequency, $f=20\mathrm{GHz}$. Based on the required wire-PEC separation distances of Fig. \ref{fig:hCondition}, which, as denoted, are frequency-invariant, we invoke Eqs. \eqref{equ:Z_tilde_condition} and \eqref{equ:capacitor_width} once more to obtain the physical dimensions of the required meta-atoms [Fig. \ref{fig:physical_configuration}(c)]. The results are given in Fig. \ref{fig:load_design_20GHz}, where we used the same procedure as before to evaluate the correction factor to be used in Eq. \eqref{equ:capacitor_width}. It was found that for $f=20\mathrm{GHz}$, this value is $K_\mathrm{corr}=0.89$, using which the predictions for the optimal $W$ [blue solid line in Fig. \ref{fig:load_design_20GHz}(b)] were obtained. As can be seen in Fig. \ref{fig:load_design_20GHz}(b), the simple relation of Eq. \eqref{equ:capacitor_width} can be still used to get good predictions for the required capacitor width at $f=20\mathrm{GHz}$. Although some of the actual optimal dimensions (red circles) deviate slightly more from the prediction, compared to the designs operating at $f=10\mathrm{GHz}$ [Fig. \ref{fig:load_design_10GHz}(b)], the deviation at these points is not very large ($\sim10\%$). Thus, the analytical relations yield a very good starting-point value, which can be readily tuned to the optimum via a short parameter sweep.

Table \ref{tab:metagrating_performance_20GHz} summarizes the design specifications and simulated scattering performance of  metagrating beam-splitters operating at $f=20\mathrm{GHz}$, corresponding to the optimal actual design points presented in Fig. \ref{fig:load_design_20GHz}(b). The field distributions are practically identical to the ones presented in Fig. \ref{fig:fields_10GHz} (not shown), with some minor differences in simulated results, stemming from different effective losses at the two frequencies. Indeed, Table \ref{tab:metagrating_performance_20GHz} reverifies that highly-effective suppression of specular reflection can be obtained via the proposed structure, corresponding to a near-unity splitting efficiency, limited only by conductor losses. Two interesting observations can be made upon comparison with the analogous designs at $f=10\mathrm{GHz}$, characterized in Fig. \ref{fig:fields_10GHz} and Table \ref{tab:metagrating_performance_10GHz}. First, losses at $f=20\mathrm{GHz}$ are smaller by $\sim20\%$ compared to the ones recorded for metagratings operating at $f=10\mathrm{GHz}$, for each of the considered split angles. Second, similar to Table \ref{tab:metagrating_performance_10GHz}, the losses are more pronounced when the split angle approaches certain working points, namely, when $\theta_\mathrm{out}\rightarrow30^{\circ}$, $\theta_\mathrm{out}\rightarrow60^{\circ}$, and $\theta_\mathrm{out}\rightarrow90^{\circ}$. 
Fortunately, the detailed analytical model presented in Section \ref{sec:theory} is highly suitable for an in-depth analysis of this intriguing loss dependency, as shall be discussed in the following subsection.

\subsection{Analysis}
\label{subsec:analysis}
Our aim in this section is to analyze the performance of the beam-splitting metagratings synthesized in Section \ref{subsec:synthesis}, when possible realistic deviations from the ideal design occur. More specifically, we would like to examine the dependency of the coupling efficiencies and Ohmic absorption in potential losses and load reactance inaccuracies, and probe the frequency response of these devices. As our detailed analytical model (Section \ref{sec:theory}) directly links the design parameters to the device performance, we utilize it to explore these relations. 

We begin by formally defining the various performance parameters to be investigated: the splitting efficiency $\eta_\mathrm{split}$ is the fraction of incident power coupled to the $\pm 1$ modes (combined); the specular reflection efficiency $\eta_\mathrm{spec}$ is the fraction coupled to specular reflection; and the losses $\eta_\mathrm{loss}$ are the fraction absorbed in the conducting wires. Decomposing the real power crossing a certain plane $z<-h$ [Eq. \eqref{equ:real_power_definition}] into the corresponding modes, identified via their spatial dependency [Eq. \eqref{equ:total_fields_below}], we can write 
\begin{equation}
\begin{array}{l}
\vspace{1mm}
\eta_\mathrm{split}=2\times\dfrac{1}{\cos\theta_\mathrm{out}}\left[\dfrac{\eta\sin\left(kh\cos\theta_\mathrm{out}\right)}{\Lambda}\right]^2
\left|\dfrac{I}{E_\mathrm{in}}\right|^2 \\
\vspace{1mm}
\eta_\mathrm{spec}=\left|1+j\dfrac{\eta\sin\left(kh\right)}{\Lambda}\dfrac{I}{E_\mathrm{in}}\right|^2 \\
\eta_\mathrm{loss}=1-\eta_\mathrm{split}-\eta_\mathrm{spec},
\end{array}
\label{equ:efficiency_definitions}
\end{equation}
where the dependency in the load impedance, not necessarily coinciding with the ideal value, enters via the fraction $I/E_\mathrm{in}$ and Ohm's law [Eq. \eqref{equ:Ohms_law_Poisson}].

Let us thus consider a general distributed load impedance $\tilde{Z}'$, not necessarily the purely-reactive optimal one $\tilde{Z}$, derived in Eq. \eqref{equ:Z_tilde_condition}. 
Thus, we can write any given load impedance as $\tilde{Z}'=\tilde{Z}+\delta\tilde{R}+j\delta\tilde{X}$, where $\delta\tilde{R}\in\mathbb{R}$ corresponds to the distributed load (conductor) resistance, responsible for losses in the system, and $\delta\tilde{X}\in\mathbb{R}$ is the deviation from the optimal distributed reactance defined by Eq. \eqref{equ:Z_tilde_condition} (e.g., due to manufacturing inaccuracies or a polychromatic excitation).

Recalling that for the devices under consideration, the wire-PEC separation $h$ satisfies Eq. \eqref{equ:h_condition}, Eq. \eqref{equ:Ohms_law_Poisson} can be inverted to yield $I/E_\mathrm{in}$ for a given (arbitrary) distributed load impedance   $\tilde{Z}'$, reading
\begin{equation}
\dfrac{I}{E_\mathrm{in}}=\dfrac{2j\sin\left(kh\right)}{\tilde{R}_g+\delta\tilde{R}+j\delta\tilde{X}},
\label{equ:I_E_fraction}
\end{equation}
where the effective grid resistance $\tilde{R}_g$ is defined as 
\begin{equation}
\tilde{R}_g=\dfrac{2\eta\sin^2\left(kh\right)}{\Lambda}=\dfrac{2\eta}{\lambda}\sin\theta_\mathrm{out}\sin^2\left(kh\right),
\label{equ:effective_grid_resistance}
\end{equation}
corresponding to the ratio between the external fields at the wire position in the absence of the wire array [Eq. \eqref{equ:external_fields}] and the current induced on the wires [Eq. \eqref{equ:specular_elimination_condition}].

Using Eq. \eqref{equ:I_E_fraction}, the coupling efficiencies of Eq. \eqref{equ:efficiency_definitions} can be explicitly written as a function of the given distributed load impedance $\tilde{Z}'$, namely,
\begin{equation}
\begin{array}{l}
\vspace{1mm}
\eta_\mathrm{split}=\dfrac{1}{\left(1+\frac{\delta\tilde{R}}{\tilde{R}_g}\right)^2+\left(\frac{\delta\tilde{X}}{\tilde{R}_g}\right)^2} \\
\vspace{1mm}
\eta_\mathrm{spec}=\dfrac{\left(\frac{\delta\tilde{R}}{\tilde{R}_g}\right)^2+\left(\frac{\delta\tilde{X}}{\tilde{R}_g}\right)^2}{\left(1+\frac{\delta\tilde{R}}{\tilde{R}_g}\right)^2+\left(\frac{\delta\tilde{X}}{\tilde{R}_g}\right)^2} \\
\eta_\mathrm{loss}=2\dfrac{\frac{\delta\tilde{R}}{\tilde{R}_g}}{\left(1+\frac{\delta\tilde{R}}{\tilde{R}_g}\right)^2+\left(\frac{\delta\tilde{X}}{\tilde{R}_g}\right)^2}.
\end{array}
\label{equ:efficiency_explicit}
\end{equation}
It can be easily verified that at the ideal optimal design point, i.e. $\delta\tilde{R}=\delta\tilde{X}=0$, the coupling efficiencies are $\eta_\mathrm{split}=1$ and $\eta_\mathrm{spec}=\eta_\mathrm{loss}=0$, in consistency with the derivation in Section \ref{sec:theory}.

\subsubsection{Conductor loss}
\label{subsubsec:losses}
To examine the effect of conductor losses on the metagrating performance, we assume that the load reactance is tuned to the optimal value ($\delta\tilde{X}=0$), and investigate the coupling efficiencies of Eq. \eqref{equ:efficiency_explicit} as a function of the load distributed resistance $\delta\tilde{R}$. It can be easily observed that the splitting efficiency gets its maximum for the lossless case $\delta\tilde{R}/\tilde{R}_g=0$, and monotonically decreases with increasing losses $\delta\tilde{R}/\tilde{R}_g>0$. 
For small losses, $\delta\tilde{R}/\tilde{R}_g\ll2$, this decrease is mainly due to the increase in absorption. Thus, the device performance deteriorates to $90\%$ of its maximal splitting efficiency approximately when $10\%$ of the incident power is lost to absorption; quantitatively, this happens when
\begin{equation}
\eta_\mathrm{loss}=10\%\Rightarrow\delta\tilde{R}_{90\%}=0.056\tilde{R}_g.
\label{equ:10_percent_loss_point}
\end{equation}

This is a very important result: it indicates that for small values of the effective grid resistance $\tilde{R}_g$, even a very small distributed wire resistance $\delta\tilde{R}$ can result in a significant amount of losses. From another perspective, for given (constant) conductor losses, the overall absorption increases \emph{inversely} proportional to $\tilde{R}_g$. In fact, for such small wire resistance $\delta\tilde{R}/\tilde{R}_g\ll1$, $\eta_\mathrm{loss}$ of Eq. \eqref{equ:efficiency_explicit} can be approximated by
\begin{equation}
\eta_\mathrm{loss}\approx2\frac{\delta\tilde{R}}{\tilde{R}_g}.
\label{equ:absorption_approximated}
\end{equation}
Thus, as revealed by Eq. \eqref{equ:effective_grid_resistance}, the working points in which the losses would be most pronounced are the ones where the product $\sin\theta_\mathrm{out}\sin^2\left(kh\right)$ is minimal, i.e. when ${h\rightarrow\nu\lambda/2},\,\nu\in\mathbb{Z}$. 
Therefore, considering the wire-PEC separation dictated by Fig. \ref{fig:hCondition}, we should expect 
increased losses at  $\theta_\mathrm{out}\rightarrow60^\circ$, where $\sin\left(kh\right)$ exactly vanishes, and around $\theta_\mathrm{out}\rightarrow30^\circ$ and $\theta_\mathrm{out}\rightarrow90^\circ$, where $\sin\left(kh\right)$ approaches zero. Indeed, this is consistent with our former observations, \textit{cf.} Tables \ref{tab:metagrating_performance_10GHz} and \ref{tab:metagrating_performance_20GHz}.

 \begin{figure}[tbh]
 \includegraphics[width=8cm]{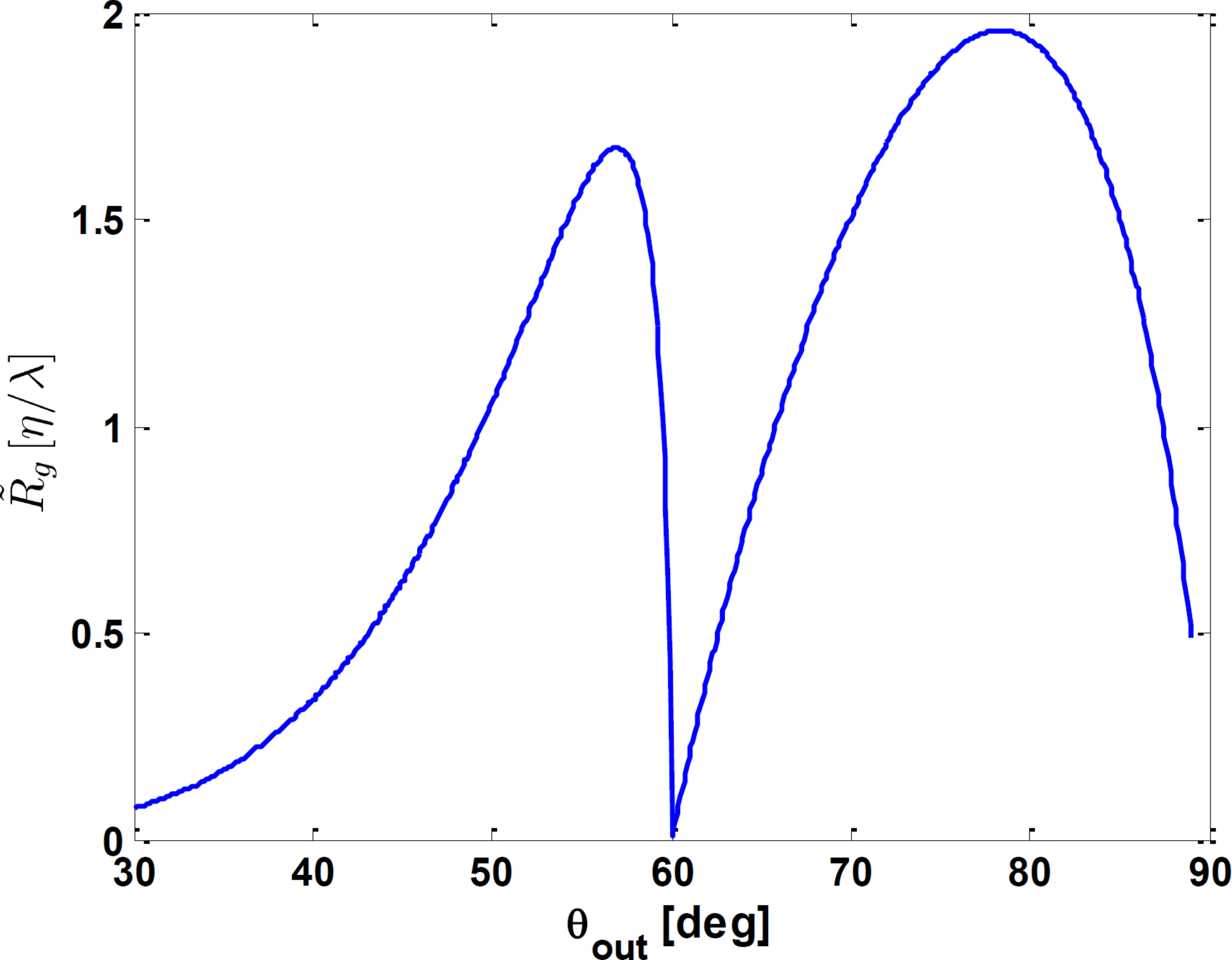}%
 \caption{Effective grid resistance as a function of the metagrating configuration corresponding to various output angles $\theta_\mathrm{out}$, following Eq. \eqref{equ:effective_grid_resistance} with $h$ of Eq. \eqref{equ:h_condition} and Fig. \ref{fig:hCondition}.} 
 \label{fig:grid_resistance}
 \end{figure}

The extent of losses, however, is not identical for all of these design points; this is due to the fact that the exact value of $\tilde{R}_g$ around its minima also depends on $\sin\theta_\mathrm{out}$, and not only on the roots of $\sin\left(kh\right)$ [Eq. \eqref{equ:effective_grid_resistance}]. This dependency is not negligible, as can be seen from Fig. \ref{fig:grid_resistance}, presenting $\tilde{R}_g$ as a function of the design parameters corresponding to $\theta_\mathrm{out}$. For a given value of $\delta\tilde{R}$, this plot predicts, for instance, that the losses approaching $\theta_\mathrm{out}=30^\circ$ will be comparable with the ones when approaching $\theta_\mathrm{out}=60^\circ$, but significantly larger than the losses very close to $\theta_\mathrm{out}=90^\circ$. 
On the other hand, Fig. \ref{fig:grid_resistance} also points out \emph{the best} working points, where the devices are the least sensitive to parasitic losses; these are indicated by the maxima of $\tilde{R}_g$, occurring around $\theta_\mathrm{out}\approx57^\circ$ and  $\theta_\mathrm{out}\approx78^\circ$. These observations, which are frequency invariant, are in consistency with the simulated results presented in Tables \ref{tab:metagrating_performance_10GHz} and \ref{tab:metagrating_performance_20GHz}.

It is not a mere coincidence that losses in these structures are inversely proportional to  $\sin\theta_\mathrm{out}\sin^2\left(kh\right)$, for a given $\delta\tilde{R}$ [Eq. \eqref{equ:absorption_approximated}]; in fact, this trend stems from a fundamental physical process taking place in these metagrating configurations. Due to interference between the current-carrying wires and their images [Eq. \eqref{equ:wire_fields_Poisson}], induced by the PEC at $z=0$, the field amplitude of the fundamental FB mode follows $\left.E_x^{\mathrm{wire}}\right|_{\mathrm{fund}}=-j\left(\eta/\lambda\right)I\sin\theta_\mathrm{out}\sin\left(kh\right)$ [Eq. \eqref{equ:total_fields_below}]. As we recall from Section \ref{subsec:specular_reflection}, this amplitude is required to meet a certain level, $E_\mathrm{in}$, in order to completely eliminate specular reflections [Eq. \eqref{equ:specular_elimination_condition}]. 

When $\sin\left(kh\right)\rightarrow0$, the phase accumulated along the distance $2kh$ is a multiple of $2\pi$; due to the $\pi$ phase shift introduced by the PEC reflection, the source and image fields tend to cancel each other at $z=-h$ [Eq. \eqref{equ:total_fields_below}]. Thus, in order to compensate this destructive interference, the design scheme tunes the metagrating configuration as to induce very large currents on the wires, to still be able to generate the fields required to eliminate specular reflection. Hence, even the slightest amount of conductor losses would result in a significant power dissipation at these working point, due to the high currents involved. On the other hand, at operating conditions for which constructive interference takes place at $z=-h$, less currents will be required, and the device would be less susceptible to losses.

Formally, we can evaluate the fraction of absorbed power as the ratio between the power dissipated per period due to induced currents flowing through resistive load and the incident power density, reading 
\begin{equation}
\eta_\mathrm{loss}=\dfrac{\frac{1}{2}\frac{\left|I\right|^2\delta\tilde{R}}{\Lambda}}{\frac{1}{2}\frac{\left|E_\mathrm{in}\right|^2}{\eta}}=\dfrac{\delta\tilde{R}}{\frac{\eta}{\lambda}\sin\theta_\mathrm{out}\sin^2\left(kh\right)}=2\frac{\delta\tilde{R}}{\tilde{R}_g},
\label{equ:absorption_from_currents}
\end{equation}
exactly as we estimated in Eq. \eqref{equ:absorption_approximated}. Indeed, the high currents developing on the wires at the points of destructive image-source interference, i.e. when the denominator is vanishing, are responsible to the observed prominent losses. Note that we have used the nominal ratio $\left|I/E_\mathrm{in}\right|$ given by Eq. \eqref{equ:specular_elimination_condition} to assess $\eta_\mathrm{loss}$ herein. For this reason, Eqs. \eqref{equ:absorption_approximated} and \eqref{equ:absorption_from_currents} are valid only for small losses $\delta\tilde{R}/\tilde{R}_g\ll1$; for more significant conductor losses, the induced current will deviate from Eq. \eqref{equ:specular_elimination_condition}, and the exact expressions Eq. \eqref{equ:efficiency_explicit} should be used.

\begin{figure}[tbh]
 \includegraphics[width=8cm]{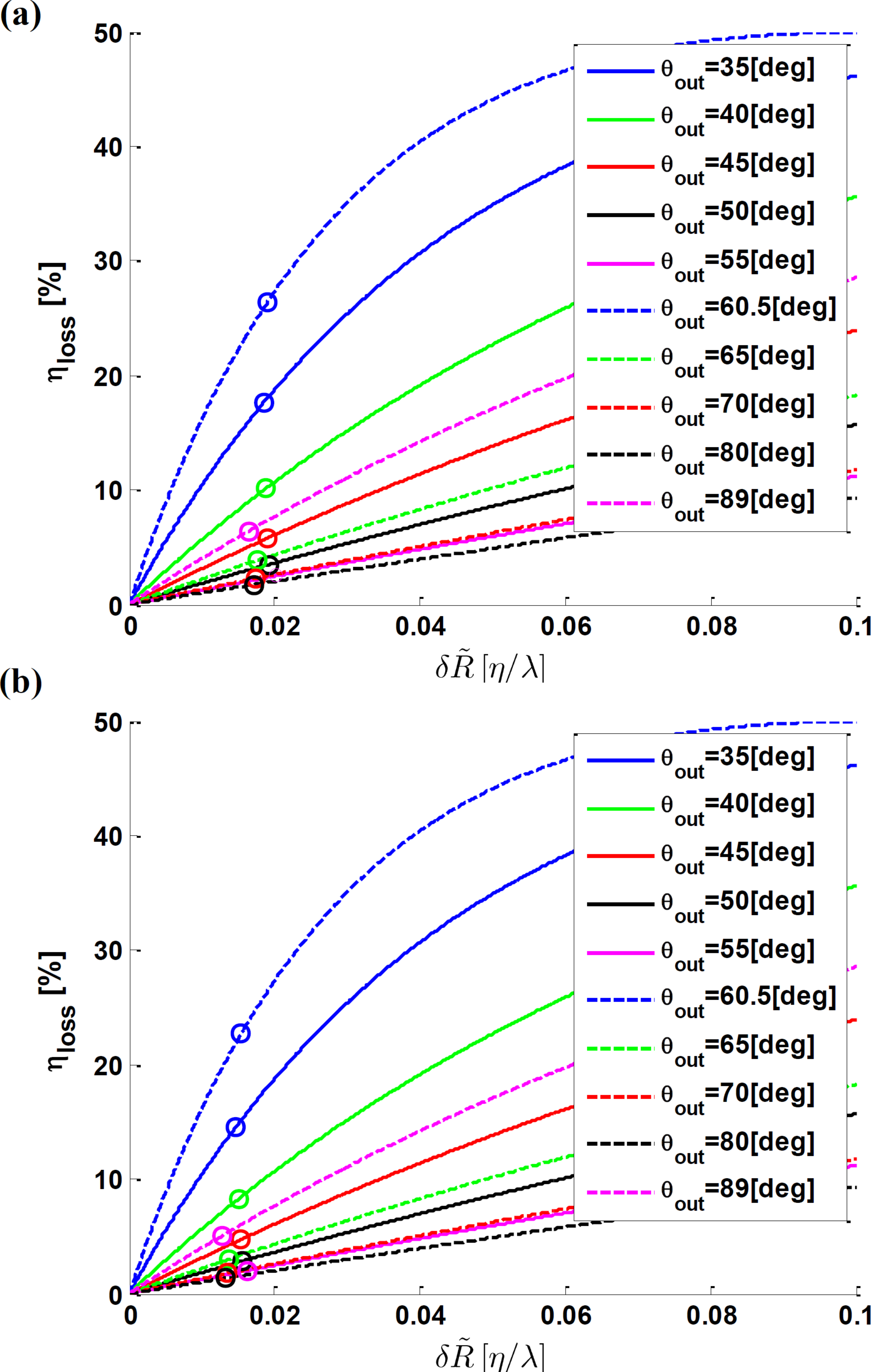}%
 \caption{Absorbed power fraction $\eta_\mathrm{loss}$ as a function of distributed conductor resistance $\delta\tilde{R}$, calculated from Eq. \eqref{equ:efficiency_explicit} for different metagrating designs, corresponding to split angles of $\theta_\mathrm{out}=35^\circ$ (blue solid line), $\theta_\mathrm{out}=40^\circ$ (green solid line), $\theta_\mathrm{out}=45^\circ$ (red solid line), $\theta_\mathrm{out}=50^\circ$ (black solid line), $\theta_\mathrm{out}=55^\circ$ (magenta solid line), $\theta_\mathrm{out}=60.5^\circ$ (blue dashed line), $\theta_\mathrm{out}=65^\circ$ (green dashed line), $\theta_\mathrm{out}=70^\circ$ (red dashed line), $\theta_\mathrm{out}=80^\circ$ (black dashed line), $\theta_\mathrm{out}=89^\circ$ (magenta dashed line). Circles denote actual losses recorded in full-wave simulations of the various designs at (a) $f=10\mathrm{GHz}$ [Table \ref{tab:metagrating_performance_10GHz}] and (b) $f=20\mathrm{GHz}$ [Table \ref{tab:metagrating_performance_20GHz}].}
 \label{fig:loss_evaluation}
 \end{figure}

Before concluding this subsection, we demonstrate how the analytical relation between $\eta_\mathrm{loss}$ and $\delta\tilde{R}$ can be harnessed to assess the distributed load resistance of the actual design. To this end, we plot in Fig. \ref{fig:loss_evaluation} the predicted absorption as a function of the distributed conductor loss $\delta\tilde{R}$, for the various metagrating beam splitters considered in Section \ref{subsec:synthesis}, calculated via Eq. \eqref{equ:efficiency_explicit}. For each considered split angle $\theta_\mathrm{out}$, corresponding to a different metagrating configuration (different $\tilde{R}_g$), we have denoted by circles the losses $\eta_\mathrm{loss}$ recorded in full-wave simulations: in Fig. \ref{fig:loss_evaluation}(a) for the $f=10\mathrm{GHz}$ metagratings, with the values documented in Table \ref{tab:metagrating_performance_10GHz}, and in Fig. \ref{fig:loss_evaluation}(b) for the $f=20\mathrm{GHz}$ metagratings, as presented in Table \ref{tab:metagrating_performance_20GHz}.

 The $\delta\tilde{R}$ values corresponding to these points represent the distributed load resistance that would, according to the theory [Eq. \eqref{equ:efficiency_explicit}], yield the observed absorption. As the conductor loss per-unit-length is mainly determined by the wire width $w$ and operating frequency (through the skin depth $\delta_\mathrm{skin}$), with a minor dependency on the capacitor width $W$, we should expect a more-or-less constant $\delta\tilde{R}$ for each one of the plots Fig. \ref{fig:loss_evaluation}(a) and (b). Indeed, Fig. \ref{fig:loss_evaluation}(a) evaluates the conductor loss at ${f=10\mathrm{GHz}}$ to be $\delta\tilde{R}=\left(18.3\pm1.2\right)\times10^{-3}\left[\eta/\lambda\right]$; at ${f=20\mathrm{GHz}}$, the values extracted from Fig. \ref{fig:loss_evaluation}(b) correspond to $\delta\tilde{R}=\left(14.5\pm1.2\right)\times10^{-3}\left[\eta/\lambda\right]$. As from Eqs. \eqref{equ:absorption_approximated} and \eqref{equ:absorption_from_currents} the absorption is approximately proportional to $\delta\tilde{R}$ for a given $\theta_\mathrm{out}$, the $\sim 20\%$ difference between the estimated $\delta\tilde{R}$ values should translate into a $\sim 20\%$ difference in $\eta_\mathrm{loss}$ at the different operating frequencies, in consistency with the results recorded in Tables \ref{tab:metagrating_performance_10GHz} and \ref{tab:metagrating_performance_20GHz}.

We compare these assessments with the analytical approximation for conductor resistance in \cite[Eq. (4.11)]{Lee2003}, treating, once more, the flat $w$-wide wire [Fig. \ref{fig:physical_configuration}(c)] as a rounded conductor with an effective radius of $r_\mathrm{eff}=w/4$ \cite{Tretyakov2003}. This results in the following approximated expression for the distributed load resistance
\begin{equation}
\delta\tilde{R}\approx\frac{1}{2\pi r_\mathrm{eff}\sigma\delta_\mathrm{skin}},
\label{equ:resistance_approximation}
\end{equation}
where the copper conductivity $\sigma$ is the same as the one used in simulations (Section \ref{subsec:synthesis}), and the skin depth is given by $\delta_\mathrm{skin}=\sqrt{2/\left(2\pi f\mu_0\sigma\right)}$; the vacuum permeability is $\mu_0=4\pi\times10^{-7}\mathrm{\left[H/m\right]}$. For the given conductor width ${w=3\mathrm{mil}=76.2\mathrm{\mu m}}$, this approximation yields ${\delta\tilde{R}=17.3\times10^{-3}\left[\eta/\lambda\right]}$ at $f=10\mathrm{GHz}$, and ${\delta\tilde{R}=12.3\times10^{-3}\left[\eta/\lambda\right]}$ at $f=20\mathrm{GHz}$, in a reasonable agreement with the average values evaluated based on Fig. \ref{fig:loss_evaluation}.

These results demonstrate the physical insight and quantitative tools provided by the detailed analytical model, directly relating the actual meta-atom geometry and constituents to the overall device losses. These relations indicate how the beam-splitter absorption can be tuned by suitable modification of the copper features, within the limitations posed by the metagrating configuration corresponding to the desirable split angle. 

\subsubsection{Reactance deviation and frequency response}
\label{subsubsec:reactance}
Next, we examine the effect of small deviations from the optimal reactance value [Eq. \eqref{equ:Z_tilde_condition}] on the metagrating performance. In terms of the expressions for the coupling efficiencies defined in Eq. \eqref{equ:efficiency_explicit}, we consider a metagrating with given (constant) conductor losses $\delta\tilde{R}$, and analyze the splitting efficiency $\eta_\mathrm{split}$ as a function of the reactance deviation $\delta\tilde{X}\neq 0$. First, we observe that, regardless of the wire resistance, the maximal splitting efficiency is achieved for $\delta\tilde{X}=0$; in other words, the value of the optimal reactance remains the one given by Eq. \eqref{equ:Z_tilde_condition}, independently of the losses in the system. This is notable, as in many devices, introduction of losses requires recalculation of the optimal reactive components (e.g., as in metasurfaces based on cascaded impedance sheets \cite{Pfeiffer2014_3}).

As before, we quantify the device sensitivity to deviation from the optimal set of parameters by calculating the reactance deviation $\delta\tilde{X}_{90\%}$ for which the splitting efficiency decreases to $90\%$ of its maximal value, for a given small distributed resistance $\delta\tilde{R}/\tilde{R}_g\ll1$. Using Eq. \eqref{equ:efficiency_explicit}, we evaluate this value as
\begin{equation}
\eta_\mathrm{split}=90\%\left.\eta_\mathrm{split}\right|_{\delta\tilde{X}=0}\Rightarrow\left|\delta\tilde{X}_{90\%}\right|\approx\!\frac{1}{3}\tilde{R}_g.
\label{equ:90_percent_splitting_point}
\end{equation}

This result indicates that the device performance is most sensitive to load reactance deviations for working points in which $\sin\theta_\mathrm{out}\sin^2\left(kh\right)$ is minimal [Eq. \eqref{equ:effective_grid_resistance}]. Although this proportionality to $\tilde{R}_g$ is very similar to the one discussed in Subsection \ref{subsubsec:losses} in the context of losses, we would like to offer here a somewhat different perspective to elucidate the origin of this dependency as it applies to reactance deviations. As discussed in the previous subsection, the wire-generated fields experience an image-source interference, affecting the ability to cancel specular reflection for a given induced current, following $\left.E_x^{\mathrm{wire}}\right|_{\mathrm{fund}}=-j\left(\eta/\lambda\right)I\sin\theta_\mathrm{out}\sin\left(kh\right)$ [Eq. \eqref{equ:total_fields_below}]. Similarly, the incident and reflected fields also undergo the same interference effects, such that the total \emph{external} field applied on the wires is ${\left.E_x^\mathrm{ext}\right|_{z=-h}=2jE_\mathrm{in}\sin\left(kh\right)}$ [Eq. \eqref{equ:external_fields}]. Effectively, this is the field that excites the current in the (passive) polarizable loaded wires, as to generate the desirable scattering phenomena.

Therefore, when $\sin\left(kh\right)\rightarrow0$, both the external fields and the wire-generated fields destructively interfere at the metagrating plane $z=-h$. In other words, for a given incident field amplitude $E_\mathrm{in}$, the external field at the metagrating plane $\left.E_x^\mathrm{ext}\right|_{z=-h}$ would be very small; thus, it would be very challenging to excite significant currents in the passive loaded wires. On the other hand, for a given induced current $I$, the amplitude of the $n=0$ FB harmonics $\left.E_x^{\mathrm{wire}}\right|_{\mathrm{fund}}$ would also be very small; thus, very high currents would be necessary to generate the fields required to eliminate specular reflection. 

Overall, around these destructive interference working points, enormous currents are generated by vanishingly-small exciting fields, by design. Consequently, the loaded wires effectively implement a transadmittance amplification system with an extremely-high gain. Therefore, any small deviation from the design specifications, equivalent to a shift in the effective "gain", would cause substantial discrepancies in the induced currents with respect to the required ones; subsequently, a rapid deterioration in the splitting efficiency is expected around these working points. According to the detailed analytical model, the severity of this double destructive interference effect can be quantified by the product of these two factors, namely, ${I/\left.E_x^\mathrm{ext}\right|_{z=-h}=1/\left[2\left(\eta/\lambda\right)\sin\theta_\mathrm{out}\sin^2\left(kh\right)\right]=1/\tilde{R}_g}$, elucidating the dependency observed in Eq. \eqref{equ:90_percent_splitting_point}.

\begin{figure}[tb]
 \includegraphics[width=8cm]{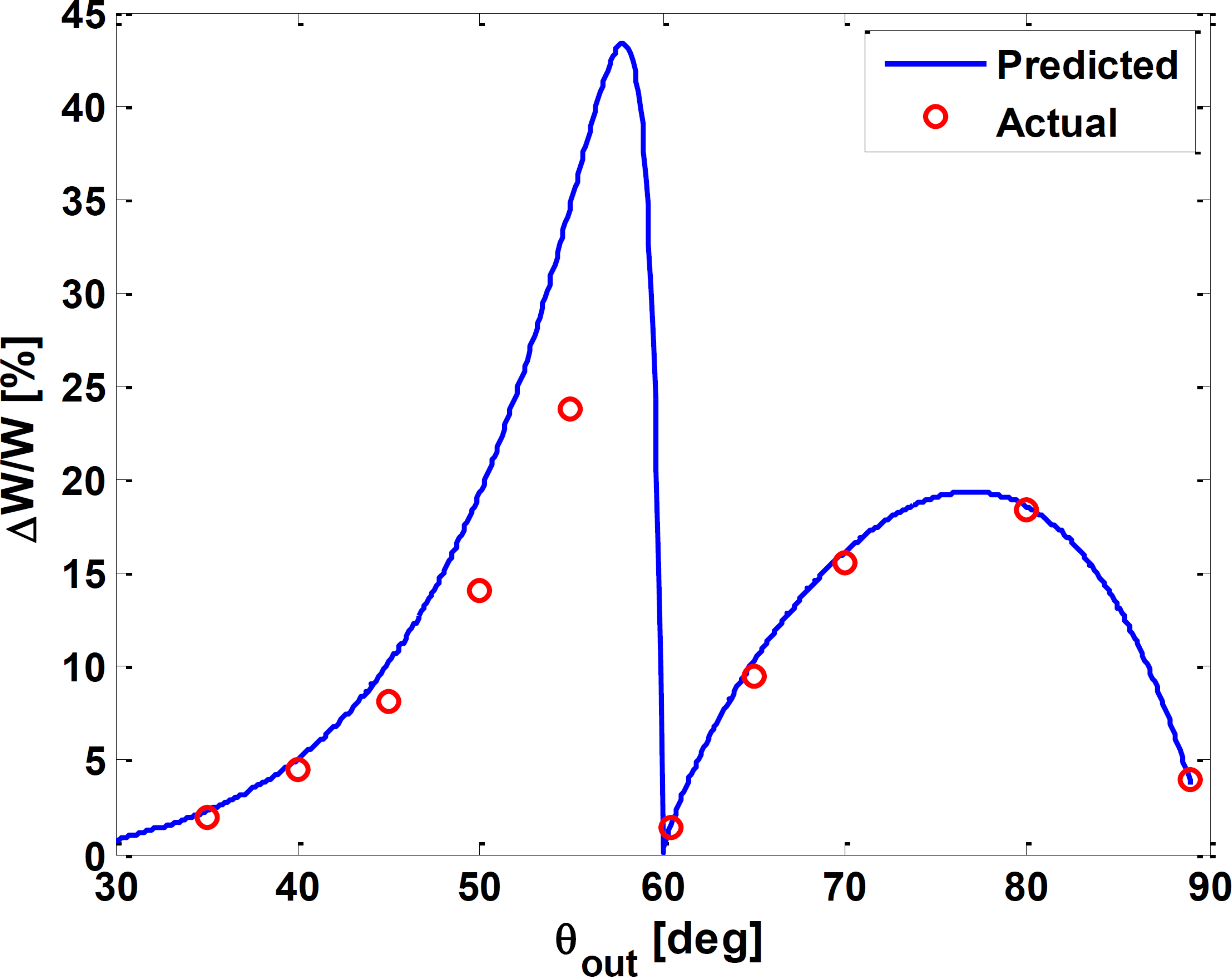}%
 \caption{Fractional capacitor width tolerance as a function of the splitting angle, for metagratings operating at $f=20\mathrm{GHz}$; the deviation range $\Delta W$ is defined as to guarantee ${\eta_\mathrm{split}\geq90\%}$. Predictions based on Eq. \eqref{equ:90_percent_splitting_point} and Eq. \eqref{equ:capacitor_width} (blue solid lines) are compared to the actual tolerances extracted from full-wave simulations of the physical structure (red circles).}
 \label{fig:capacitor_width_tolerance_20GHz}
 \end{figure}

We can use Eq. \eqref{equ:90_percent_splitting_point} in conjunction with Eq. \eqref{equ:capacitor_width} to estimate the maximal allowed deviation in the capacitor width that would still retain $\eta_\mathrm{split}$ above $90\%$ of its maximum. The fractional capacitor-width deviation tolerance, $\Delta W/W$, predicted correspondingly, is presented in Fig. \ref{fig:capacitor_width_tolerance_20GHz} as a function of the split angle, for the metagratings synthesized in Section \ref{subsec:synthesis}; for brevity, results are shown only for the designs operating at $f=20\mathrm{GHz}$. Simultaneously, we have extracted from full-wave simulations the actual tolerances obtained for the corresponding physical realizations [Fig. \ref{fig:physical_configuration}(c)]; these are denoted as red circles in Fig. \ref{fig:capacitor_width_tolerance_20GHz}. The good agreement between the predicted and simulated values serves as another verification of the analytical model, demonstrating its efficacy in assessing the performance of a given design in terms of the detailed meta-atom geometrical parameters. Note that the working points in which slightly larger discrepancies occur are the ones for which the analytical model incurs slight errors in predicting the optimal capacitor width to begin with [Fig. \ref{fig:load_design_20GHz}(b)]. 

A comparison between Fig. \ref{fig:capacitor_width_tolerance_20GHz} and Fig. \ref{fig:grid_resistance} indicates that, as implied by Eq. \eqref{equ:90_percent_splitting_point}, the tolerance to inaccuracies in the load reactance follows closely the trend of $\tilde{R}_g$. Specifically, the most sensitive working points occur for $\theta_\mathrm{out}\rightarrow30^\circ$, $\theta_\mathrm{out}\rightarrow60^\circ$, and $\theta_\mathrm{out}\rightarrow90^\circ$, where $\tilde{R}_g$ approaches its minima, and the highest tolerance is recorded around $\theta_\mathrm{out}\approx58^\circ$ and $\theta_\mathrm{out}\approx77^\circ$, very close to the maxima of $\tilde{R}_g$. Nevertheless, a closer examination reveals that the position of the \emph{global} maximum in the two figures is different. This is due to the fact that the fractional capacitor-width tolerance is dependent also at the nominal value of $W$, corresponding to the nominal reactance at each of the working points (Fig. \ref{fig:load_design_20GHz}); however, these nominal values are not taken into account in Eq. \eqref{equ:90_percent_splitting_point}. Therefore, while the general trends should be very similar, some quantitative differences are expected.

The same physical considerations lead us to hypothesize that the tolerance to changes in the operating frequency should also follow a trend similar to that of $\tilde{R}_g$. As discussed after Eq. \eqref{equ:90_percent_splitting_point}, at the points where the double destructive interference occur, the metagrating exhibits an extreme sensitivity to deviations from the nominal design parameters, due to the astronomical by-design induced-current-to-applied-field ratio. Correspondingly, around these working points we would expect the smallest operational bandwidth. Evaluating the $90\%$ splitting efficiency bandwidth in closed form is more complicated, as frequency variations modify the effective splitting angle following Eq. \eqref{equ:Lambda_period}, as well as cause deviations from the relation Eq. \eqref{equ:Z_tilde_condition} between the load impedance and metagrating geometry
; while linearization of the frequency response is possible, the analytical expressions are cumbersome, and yield little physical intuition. On the other hand, the bandwidth can be implicitly evaluated from the analytical model in a straightforward manner, allowing us to probe our hypothesis.

To this end, we calculate the scattered fields for metagratings designed at $f=20\mathrm{GHz}$ (i.e. with fixed $h$, $W$, and $\Lambda$, extracted, respectively, from Fig. \ref{fig:hCondition}, Fig. \ref{fig:load_design_20GHz}, and Eq. \eqref{equ:Lambda_period}), excited by normally-incident plane waves at different frequencies. As the distributed reactance at $f=20\mathrm{GHz}$ is known and is capacitive [Fig. \ref{fig:load_design_20GHz}(a)], the load reactance as a function of frequency can be readily deduced by considering the typical inverse proportional dependency in frequency. Hence, the problem at hand reduces to the one of scattering off a \emph{given} loaded wire array in front of a PEC, for which the fields below the metagrating are given by Eq. \eqref{equ:total_fields_below}, with the induced current $I$ evaluated via Eq. \eqref{equ:Ohms_law_Poisson}. The fraction of the incident power coupled to the various FB modes can be subsequently assessed from Eq. \eqref{equ:efficiency_definitions}. Note that when deriving these equations, we did not assume anything regarding the values of the metagrating parameters, making them applicable for the desirable calculation.

\begin{figure}[tb]
 \includegraphics[width=8cm]{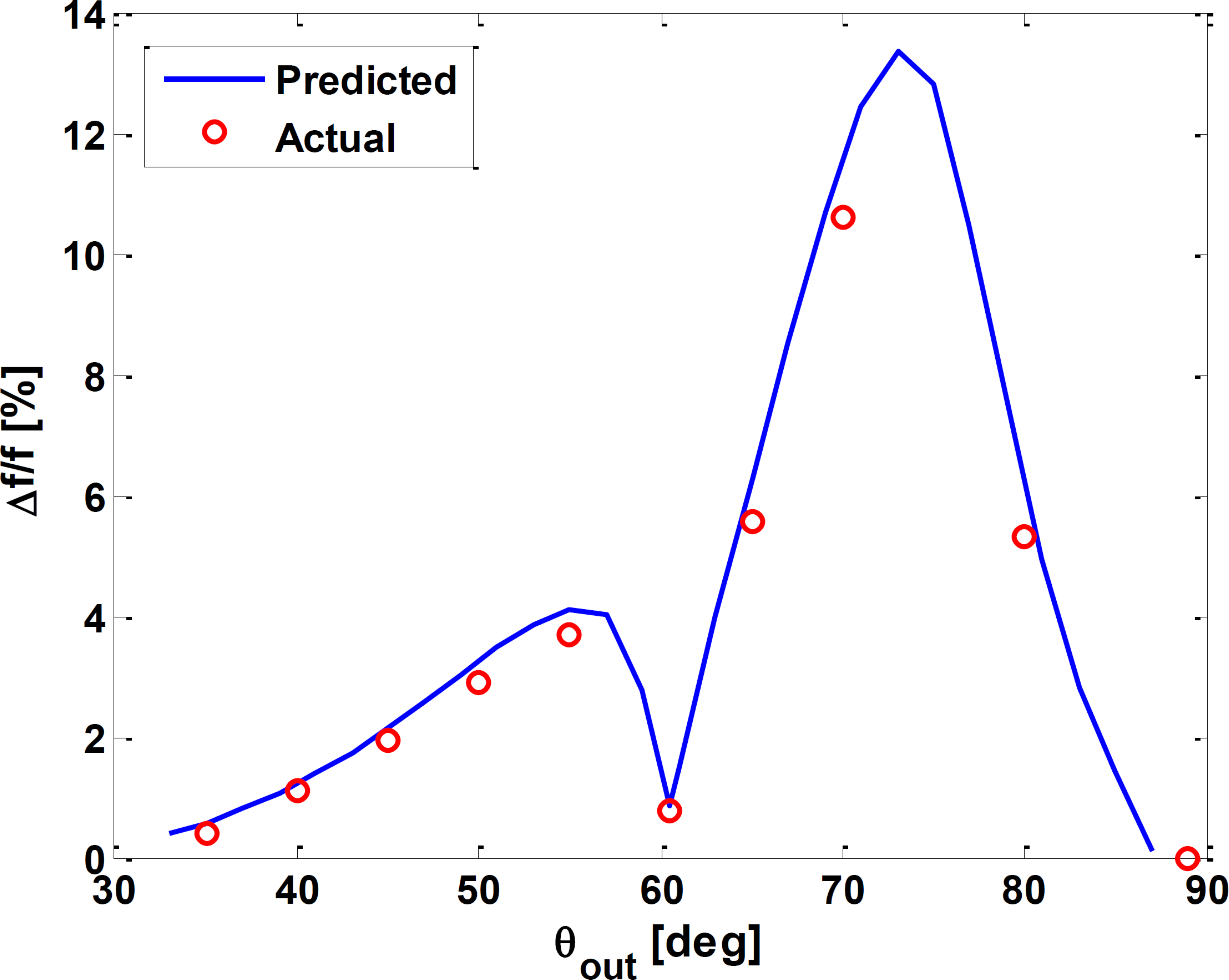}%
 \caption{Fractional frequency bandwidth as a function of the splitting angle, for metagratings designed for operation at $f=20\mathrm{GHz}$; the deviation range $\Delta f$ is defined as to guarantee ${\eta_\mathrm{split}\geq90\%}$. Predictions based on the analytical model (blue solid lines) are compared to the actual bandwidth extracted from full-wave simulations of the physical structure (red circles).}
 \label{fig:frequency_bandwidth_20GHz}
 \end{figure}

The fractional $90\%$ splitting-efficiency bandwidth calculated correspondingly from the analytical model is presented in Fig. \ref{fig:frequency_bandwidth_20GHz} (blue solid line), along with the bandwidths extracted from the simulated metagrating geometries (red circles), as a function of the various split angles. The predicted and actual frequency bandwidths agree remarkably, demonstrating the high accuracy of the formulation when applied to realistic physical structures. 

We note that within the frequency range indicated by $\Delta f$ the splitting efficiency remains very high, although the actual split angle varies with frequency [Eq. \eqref{equ:Lambda_period}]. Towards the edges of the split-angle interval $\left(30^\circ,90^\circ\right)$, frequency changes may drive the $\pm1$ FB modes towards the evanescent spectrum, or allow higher FB modes to be excited, which also limits the achievable bandwidths. These bandwidths may not seem very impressive at first sight; however, one should bear in mind that these refer to $90\%$ performance bandwidths, and not to the typical $50\%$ (or 3dB) performance points. Hence, the values plotted in Fig. \ref{fig:frequency_bandwidth_20GHz} actually correspond to a rather moderate frequency response (at least away from the plot minima), in consistency with the observations of \cite{Radi2017}.

Importantly, the evaluated fractional bandwidths confirm our hypothesis, as their trend clearly follows the one of the effective grid resistance [Fig. \ref{fig:grid_resistance}]. Indeed, the working points in which the image-source interference causes high currents to be induced in response to very small applied fields ($\theta_\mathrm{out}\rightarrow30^\circ$, $\theta_\mathrm{out}\rightarrow60^\circ$, and $\theta_\mathrm{out}\rightarrow90^\circ$) exhibit the smallest bandwidths, due to the high sensitivity to small variations in the design parameters [see discussion after Eq. \eqref{equ:90_percent_splitting_point}]. On the other hand, away from these points of destructive interference, the device performance is quite stable with respect to moderate frequency variations, up to the inevitable change in the split angle.

\section{Conclusion}
\label{sec:conclusion}
To conclude, we have presented a detailed analytical model for metagrating beam splitters, based on loaded conducting wire arrays. With respect to previous reports, the formulation describes electrically-polarizable metagratings excited by TE-polarized fields, more practical for realization of planar devices, and derives explicit relations between the device performance parameters and the individual meta-atom load, including realistic losses. From a synthesis perspective, these relations allow an almost-analytical prediction of the required meta-atom geometry, significantly reducing the design effort. From an analysis point of view, the ability to naturally integrate conductor losses, and deviations from the nominal reactance and frequency operating conditions, provide a convenient analytical framework to investigate the effects of these parasitics on the metagrating performance.

Specifically, we have revealed that the metagratings feature distinct preferable working points. 
Both in terms of losses, as well in terms of reactance deviation and frequency response, designs that operate close to the points where the effective grid resistance $\tilde{R}_g$ tends to zero are more prone to significant performance reduction, exhibiting extremely high sensitivity to conductor losses, load geometry inaccuracies, and frequency shifts. Relying on the analytical derivation, we have shown that these phenomena stem from fundamental interference processes taking place in the device. At these wire-PEC separation distances where destructive interference occurs for both the incident and wire-generated fields, extremely-high currents are expected to be excited by overall extremely-low effective fields. These extreme operating conditions lead to high sensitivity to design parameters as well as to significant losses, due to the large by-design transadmittance "gain" and large conducted currents. These physical effects are very basic and general, and thus are expected to be observed in any metagrating system of this sort. 

Interestingly, these problematic working points are not correlated with the typical challenging operating conditions of beam-manipulating metasurfaces \cite{Epstein2014, Epstein2014_2, Wong2016, Epstein2016_3, Epstein2016_4, Asadchy2016}, in which performance reduction is commonly associated with large wave-impedance mismatch. In fact, for the investigated metagrating devices, some of the best working points actually occur for extremely wide-angle beam splitting.

The detailed model, verified with full-wave simulations of realistic physical structures, thus provides both a set of efficient semianalytical tools for synthesis and analysis, and physical insight regarding the dominant processes taking place within the device. Our observations also highlight the immense potential of these devices for a variety of wave-manipulating devices, in consistency with previous reports \cite{Sounas2016, Wong2017, Memarian2017, PaniaguaDominguez2017, Radi2017, Wong2017_1}. In particular, when suitable working points are chosen, these metagratings can split a normally-incident beam into two equal-power beams propagating at very large oblique angles ($\sim80^\circ$) with minimal absorption, moderate bandwidth, and substantial resilience to fabrication inaccuracies. In fact, such a perfect wide-angle reflect-mode beam-splitting is still considered a very challenging problem to solve accurately with conventional metasurfaces \cite{Estakhri2016,Estakhri2016_1}, even though metagratings feature a much simpler structure, requiring only the design of a single meta-atom (which can be done semianalytically following our derivation).

Finally, it is important to note that although the synthesis and analysis presented herein were demonstrated using metagratings operating at microwave frequencies, the derivation and observations are not restricted to this frequency range. More than that, the same meta-atom structures have been used in the past to devise metasurfaces for terahertz and optical applications \cite{Zhao2012, Pfeiffer2014, Kuznetsov2015, Chang2017}. Hence, the presented analytical model could facilitate effective semianalytical design of novel low-loss, robust, ultrathin devices for field manipulation across the electromagnetic spectrum, with the highlighted physical observations guiding the synthesis to enhance performance by judicious choice of working points.

\bibliography{metaGratingAnalyticalModel}

\begin{thebibliography}{63}%
\makeatletter
\providecommand \@ifxundefined [1]{%
 \@ifx{#1\undefined}
}%
\providecommand \@ifnum [1]{%
 \ifnum #1\expandafter \@firstoftwo
 \else \expandafter \@secondoftwo
 \fi
}%
\providecommand \@ifx [1]{%
 \ifx #1\expandafter \@firstoftwo
 \else \expandafter \@secondoftwo
 \fi
}%
\providecommand \natexlab [1]{#1}%
\providecommand \enquote  [1]{``#1''}%
\providecommand \bibnamefont  [1]{#1}%
\providecommand \bibfnamefont [1]{#1}%
\providecommand \citenamefont [1]{#1}%
\providecommand \href@noop [0]{\@secondoftwo}%
\providecommand \href [0]{\begingroup \@sanitize@url \@href}%
\providecommand \@href[1]{\@@startlink{#1}\@@href}%
\providecommand \@@href[1]{\endgroup#1\@@endlink}%
\providecommand \@sanitize@url [0]{\catcode `\\12\catcode `\$12\catcode
  `\&12\catcode `\#12\catcode `\^12\catcode `\_12\catcode `\%12\relax}%
\providecommand \@@startlink[1]{}%
\providecommand \@@endlink[0]{}%
\providecommand \url  [0]{\begingroup\@sanitize@url \@url }%
\providecommand \@url [1]{\endgroup\@href {#1}{\urlprefix }}%
\providecommand \urlprefix  [0]{URL }%
\providecommand \Eprint [0]{\href }%
\providecommand \doibase [0]{http://dx.doi.org/}%
\providecommand \selectlanguage [0]{\@gobble}%
\providecommand \bibinfo  [0]{\@secondoftwo}%
\providecommand \bibfield  [0]{\@secondoftwo}%
\providecommand \translation [1]{[#1]}%
\providecommand \BibitemOpen [0]{}%
\providecommand \bibitemStop [0]{}%
\providecommand \bibitemNoStop [0]{.\EOS\space}%
\providecommand \EOS [0]{\spacefactor3000\relax}%
\providecommand \BibitemShut  [1]{\csname bibitem#1\endcsname}%
\let\auto@bib@innerbib\@empty
\bibitem [{\citenamefont {Pfeiffer}\ and\ \citenamefont
  {Grbic}(2013)}]{Pfeiffer2013}%
  \BibitemOpen
  \bibfield  {author} {\bibinfo {author} {\bibfnamefont {C.}~\bibnamefont
  {Pfeiffer}}\ and\ \bibinfo {author} {\bibfnamefont {A.}~\bibnamefont
  {Grbic}},\ }\bibfield  {title} {\enquote {\bibinfo {title} {{Metamaterial
  Huygens' surfaces: tailoring wave fronts with reflectionless sheets}},}\
  }\href {\doibase 10.1103/PhysRevLett.110.197401} {\bibfield  {journal}
  {\bibinfo  {journal} {Phys. Rev. Lett.}\ }\textbf {\bibinfo {volume} {110}},\
  \bibinfo {pages} {197401} (\bibinfo {year} {2013})}\BibitemShut {NoStop}%
\bibitem [{\citenamefont {Monticone}\ \emph {et~al.}(2013)\citenamefont
  {Monticone}, \citenamefont {Estakhri},\ and\ \citenamefont
  {Al\`{u}}}]{Monticone2013}%
  \BibitemOpen
  \bibfield  {author} {\bibinfo {author} {\bibfnamefont {F.}~\bibnamefont
  {Monticone}}, \bibinfo {author} {\bibfnamefont {N.~M.}\ \bibnamefont
  {Estakhri}}, \ and\ \bibinfo {author} {\bibfnamefont {A.}~\bibnamefont
  {Al\`{u}}},\ }\bibfield  {title} {\enquote {\bibinfo {title} {Full control of
  nanoscale optical transmission with a composite metascreen},}\ }\href@noop {}
  {\bibfield  {journal} {\bibinfo  {journal} {Phys. Rev. Lett.}\ }\textbf
  {\bibinfo {volume} {110}},\ \bibinfo {pages} {203903} (\bibinfo {year}
  {2013})}\BibitemShut {NoStop}%
\bibitem [{\citenamefont {Selvanayagam}\ and\ \citenamefont
  {Eleftheriades}(2013)}]{Selvanayagam2013}%
  \BibitemOpen
  \bibfield  {author} {\bibinfo {author} {\bibfnamefont {M.}~\bibnamefont
  {Selvanayagam}}\ and\ \bibinfo {author} {\bibfnamefont {G.~V.}\ \bibnamefont
  {Eleftheriades}},\ }\bibfield  {title} {\enquote {\bibinfo {title}
  {Discontinuous electromagnetic fields using orthogonal electric and magnetic
  currents for wavefront manipulation},}\ }\href {\doibase
  10.1364/OE.21.014409} {\bibfield  {journal} {\bibinfo  {journal} {Opt.
  Express}\ }\textbf {\bibinfo {volume} {21}},\ \bibinfo {pages} {14409--14429}
  (\bibinfo {year} {2013})}\BibitemShut {NoStop}%
\bibitem [{\citenamefont {Epstein}\ and\ \citenamefont
  {Eleftheriades}(2016{\natexlab{a}})}]{Epstein2016_3}%
  \BibitemOpen
  \bibfield  {author} {\bibinfo {author} {\bibfnamefont {A.}~\bibnamefont
  {Epstein}}\ and\ \bibinfo {author} {\bibfnamefont {G.~V.}\ \bibnamefont
  {Eleftheriades}},\ }\bibfield  {title} {\enquote {\bibinfo {title} {Arbitrary
  power-conserving field transformations with passive lossless omega-type
  bianisotropic metasurfaces},}\ }\href@noop {} {\bibfield  {journal} {\bibinfo
   {journal} {IEEE Trans. Antennas Propag.}\ }\textbf {\bibinfo {volume}
  {64}},\ \bibinfo {pages} {3880--3895} (\bibinfo {year}
  {2016}{\natexlab{a}})}\BibitemShut {NoStop}%
\bibitem [{\citenamefont {Cui}\ \emph {et~al.}(2014)\citenamefont {Cui},
  \citenamefont {Qi}, \citenamefont {Wan}, \citenamefont {Zhao},\ and\
  \citenamefont {Cheng}}]{Cui2014}%
  \BibitemOpen
  \bibfield  {author} {\bibinfo {author} {\bibfnamefont {T.~J.}\ \bibnamefont
  {Cui}}, \bibinfo {author} {\bibfnamefont {M.~Q.}\ \bibnamefont {Qi}},
  \bibinfo {author} {\bibfnamefont {X.}~\bibnamefont {Wan}}, \bibinfo {author}
  {\bibfnamefont {J.}~\bibnamefont {Zhao}}, \ and\ \bibinfo {author}
  {\bibfnamefont {Q.}~\bibnamefont {Cheng}},\ }\bibfield  {title} {\enquote
  {\bibinfo {title} {{Coding Metamaterials, Digital Metamaterials and
  Programming Metamaterials}},}\ }\href {\doibase 10.1038/lsa.2014.99}
  {\bibfield  {journal} {\bibinfo  {journal} {Light Sci. Appl.}\ }\textbf
  {\bibinfo {volume} {3}},\ \bibinfo {pages} {19} (\bibinfo {year}
  {2014})}\BibitemShut {NoStop}%
\bibitem [{\citenamefont {Asadchy}\ \emph
  {et~al.}(2015{\natexlab{a}})\citenamefont {Asadchy}, \citenamefont {Ra'di},
  \citenamefont {Vehmas},\ and\ \citenamefont {Tretyakov}}]{Asadchy2015}%
  \BibitemOpen
  \bibfield  {author} {\bibinfo {author} {\bibfnamefont {V.~S.}\ \bibnamefont
  {Asadchy}}, \bibinfo {author} {\bibfnamefont {Y.}~\bibnamefont {Ra'di}},
  \bibinfo {author} {\bibfnamefont {J.}~\bibnamefont {Vehmas}}, \ and\ \bibinfo
  {author} {\bibfnamefont {S.~A.}\ \bibnamefont {Tretyakov}},\ }\bibfield
  {title} {\enquote {\bibinfo {title} {Functional metamirrors using
  bianisotropic elements},}\ }\href@noop {} {\bibfield  {journal} {\bibinfo
  {journal} {Phys. Rev. Lett.}\ }\textbf {\bibinfo {volume} {114}},\ \bibinfo
  {pages} {095503} (\bibinfo {year} {2015}{\natexlab{a}})}\BibitemShut
  {NoStop}%
\bibitem [{\citenamefont {Asadchy}\ \emph
  {et~al.}(2016{\natexlab{a}})\citenamefont {Asadchy}, \citenamefont
  {Albooyeh}, \citenamefont {Tcvetkova}, \citenamefont {D\'{\i}az-Rubio},
  \citenamefont {Ra'di},\ and\ \citenamefont {Tretyakov}}]{Asadchy2016}%
  \BibitemOpen
  \bibfield  {author} {\bibinfo {author} {\bibfnamefont {V.~S.}\ \bibnamefont
  {Asadchy}}, \bibinfo {author} {\bibfnamefont {M.}~\bibnamefont {Albooyeh}},
  \bibinfo {author} {\bibfnamefont {S.~N.}\ \bibnamefont {Tcvetkova}}, \bibinfo
  {author} {\bibfnamefont {A.}~\bibnamefont {D\'{\i}az-Rubio}}, \bibinfo
  {author} {\bibfnamefont {Y.}~\bibnamefont {Ra'di}}, \ and\ \bibinfo {author}
  {\bibfnamefont {S.~A.}\ \bibnamefont {Tretyakov}},\ }\bibfield  {title}
  {\enquote {\bibinfo {title} {Perfect control of reflection and refraction
  using spatially dispersive metasurfaces},}\ }\href {\doibase
  10.1103/PhysRevB.94.075142} {\bibfield  {journal} {\bibinfo  {journal} {Phys.
  Rev. B}\ }\textbf {\bibinfo {volume} {94}},\ \bibinfo {pages} {075142}
  (\bibinfo {year} {2016}{\natexlab{a}})}\BibitemShut {NoStop}%
\bibitem [{\citenamefont {{Mohammadi Estakhri}}\ and\ \citenamefont
  {Al{\`{u}}}(2016)}]{Estakhri2016_1}%
  \BibitemOpen
  \bibfield  {author} {\bibinfo {author} {\bibfnamefont {N.}~\bibnamefont
  {{Mohammadi Estakhri}}}\ and\ \bibinfo {author} {\bibfnamefont
  {A.}~\bibnamefont {Al{\`{u}}}},\ }\bibfield  {title} {\enquote {\bibinfo
  {title} {Wave-front transformation with gradient metasurfaces},}\ }\href
  {\doibase 10.1103/PhysRevX.6.041008} {\bibfield  {journal} {\bibinfo
  {journal} {Phys. Rev. X}\ }\textbf {\bibinfo {volume} {6}},\ \bibinfo {pages}
  {041008} (\bibinfo {year} {2016})}\BibitemShut {NoStop}%
\bibitem [{\citenamefont {Epstein}\ and\ \citenamefont
  {Eleftheriades}(2016{\natexlab{b}})}]{Epstein2016_4}%
  \BibitemOpen
  \bibfield  {author} {\bibinfo {author} {\bibfnamefont {A.}~\bibnamefont
  {Epstein}}\ and\ \bibinfo {author} {\bibfnamefont {G.~V.}\ \bibnamefont
  {Eleftheriades}},\ }\bibfield  {title} {\enquote {\bibinfo {title} {Synthesis
  of passive lossless metasurfaces using auxiliary fields for reflectionless
  beam splitting and perfect reflection},}\ }\href@noop {} {\bibfield
  {journal} {\bibinfo  {journal} {Phys. Rev. Lett.}\ }\textbf {\bibinfo
  {volume} {117}},\ \bibinfo {pages} {256103} (\bibinfo {year}
  {2016}{\natexlab{b}})}\BibitemShut {NoStop}%
\bibitem [{\citenamefont {Lin}\ \emph {et~al.}(2014)\citenamefont {Lin},
  \citenamefont {Fan}, \citenamefont {Hasman},\ and\ \citenamefont
  {Brongersma}}]{Lin2014}%
  \BibitemOpen
  \bibfield  {author} {\bibinfo {author} {\bibfnamefont {D.}~\bibnamefont
  {Lin}}, \bibinfo {author} {\bibfnamefont {P.}~\bibnamefont {Fan}}, \bibinfo
  {author} {\bibfnamefont {E.}~\bibnamefont {Hasman}}, \ and\ \bibinfo {author}
  {\bibfnamefont {M.~L.}\ \bibnamefont {Brongersma}},\ }\bibfield  {title}
  {\enquote {\bibinfo {title} {Dielectric gradient metasurface optical
  elements},}\ }\href {\doibase 10.1126/science.1253213} {\bibfield  {journal}
  {\bibinfo  {journal} {Science}\ }\textbf {\bibinfo {volume} {345}},\ \bibinfo
  {pages} {298--302} (\bibinfo {year} {2014})}\BibitemShut {NoStop}%
\bibitem [{\citenamefont {Aieta}\ \emph {et~al.}(2015)\citenamefont {Aieta},
  \citenamefont {Kats}, \citenamefont {Genevet},\ and\ \citenamefont
  {Capasso}}]{Aieta2015}%
  \BibitemOpen
  \bibfield  {author} {\bibinfo {author} {\bibfnamefont {F.}~\bibnamefont
  {Aieta}}, \bibinfo {author} {\bibfnamefont {M.~A.}\ \bibnamefont {Kats}},
  \bibinfo {author} {\bibfnamefont {P.}~\bibnamefont {Genevet}}, \ and\
  \bibinfo {author} {\bibfnamefont {F.}~\bibnamefont {Capasso}},\ }\bibfield
  {title} {\enquote {\bibinfo {title} {Multiwavelength achromatic metasurfaces
  by dispersive phase compensation},}\ }\href {\doibase
  10.1126/science.aaa2494} {\bibfield  {journal} {\bibinfo  {journal}
  {Science}\ }\textbf {\bibinfo {volume} {347}},\ \bibinfo {pages} {1342--1345}
  (\bibinfo {year} {2015})}\BibitemShut {NoStop}%
\bibitem [{\citenamefont {Zhao}\ \emph {et~al.}(2012)\citenamefont {Zhao},
  \citenamefont {Belkin},\ and\ \citenamefont {Al\`{u}}}]{Zhao2012}%
  \BibitemOpen
  \bibfield  {author} {\bibinfo {author} {\bibfnamefont {Y.}~\bibnamefont
  {Zhao}}, \bibinfo {author} {\bibfnamefont {M.~A.}\ \bibnamefont {Belkin}}, \
  and\ \bibinfo {author} {\bibfnamefont {A.}~\bibnamefont {Al\`{u}}},\
  }\bibfield  {title} {\enquote {\bibinfo {title} {{Twisted optical
  metamaterials for planarized ultrathin broadband circular polarizers.}}}\
  }\href {\doibase 10.1038/ncomms1877} {\bibfield  {journal} {\bibinfo
  {journal} {Nat. Commun.}\ }\textbf {\bibinfo {volume} {3}},\ \bibinfo {pages}
  {870} (\bibinfo {year} {2012})}\BibitemShut {NoStop}%
\bibitem [{\citenamefont {Pfeiffer}\ \emph
  {et~al.}(2014{\natexlab{a}})\citenamefont {Pfeiffer}, \citenamefont {Zhang},
  \citenamefont {Ray}, \citenamefont {Guo},\ and\ \citenamefont
  {Grbic}}]{Pfeiffer2014_1}%
  \BibitemOpen
  \bibfield  {author} {\bibinfo {author} {\bibfnamefont {C.}~\bibnamefont
  {Pfeiffer}}, \bibinfo {author} {\bibfnamefont {C.}~\bibnamefont {Zhang}},
  \bibinfo {author} {\bibfnamefont {V.}~\bibnamefont {Ray}}, \bibinfo {author}
  {\bibfnamefont {L.~J.}\ \bibnamefont {Guo}}, \ and\ \bibinfo {author}
  {\bibfnamefont {A.}~\bibnamefont {Grbic}},\ }\bibfield  {title} {\enquote
  {\bibinfo {title} {High performance bianisotropic metasurfaces: Asymmetric
  transmission of light},}\ }\href@noop {} {\bibfield  {journal} {\bibinfo
  {journal} {Phys. Rev. Lett.}\ }\textbf {\bibinfo {volume} {113}},\ \bibinfo
  {pages} {023902} (\bibinfo {year} {2014}{\natexlab{a}})}\BibitemShut
  {NoStop}%
\bibitem [{\citenamefont {Pfeiffer}\ and\ \citenamefont
  {Grbic}(2014)}]{Pfeiffer2014_3}%
  \BibitemOpen
  \bibfield  {author} {\bibinfo {author} {\bibfnamefont {C.}~\bibnamefont
  {Pfeiffer}}\ and\ \bibinfo {author} {\bibfnamefont {A.}~\bibnamefont
  {Grbic}},\ }\bibfield  {title} {\enquote {\bibinfo {title} {Bianisotropic
  metasurfaces for optimal polarization control: Analysis and synthesis},}\
  }\href {\doibase 10.1103/PhysRevApplied.2.044011} {\bibfield  {journal}
  {\bibinfo  {journal} {Phys. Rev. Appl.}\ }\textbf {\bibinfo {volume} {2}},\
  \bibinfo {pages} {044011} (\bibinfo {year} {2014})}\BibitemShut {NoStop}%
\bibitem [{\citenamefont {Achouri}\ \emph {et~al.}(2015)\citenamefont
  {Achouri}, \citenamefont {Khan}, \citenamefont {Gupta}, \citenamefont
  {Lavigne}, \citenamefont {Salem},\ and\ \citenamefont
  {Caloz}}]{Achouri2015_1}%
  \BibitemOpen
  \bibfield  {author} {\bibinfo {author} {\bibfnamefont {K.}~\bibnamefont
  {Achouri}}, \bibinfo {author} {\bibfnamefont {B.~A.}\ \bibnamefont {Khan}},
  \bibinfo {author} {\bibfnamefont {S.}~\bibnamefont {Gupta}}, \bibinfo
  {author} {\bibfnamefont {G.}~\bibnamefont {Lavigne}}, \bibinfo {author}
  {\bibfnamefont {M.~A.}\ \bibnamefont {Salem}}, \ and\ \bibinfo {author}
  {\bibfnamefont {C.}~\bibnamefont {Caloz}},\ }\bibfield  {title} {\enquote
  {\bibinfo {title} {Synthesis of electromagnetic metasurfaces: principles and
  illustrations},}\ }\href {\doibase 10.1051/epjam/2015016} {\bibfield
  {journal} {\bibinfo  {journal} {EPJ Appl. Metamaterials}\ }\textbf {\bibinfo
  {volume} {2}},\ \bibinfo {pages} {12} (\bibinfo {year} {2015})}\BibitemShut
  {NoStop}%
\bibitem [{\citenamefont {Yin}\ \emph {et~al.}(2015)\citenamefont {Yin},
  \citenamefont {Wan}, \citenamefont {Zhang},\ and\ \citenamefont
  {Cui}}]{Yin2015}%
  \BibitemOpen
  \bibfield  {author} {\bibinfo {author} {\bibfnamefont {J.~Y.}\ \bibnamefont
  {Yin}}, \bibinfo {author} {\bibfnamefont {X.}~\bibnamefont {Wan}}, \bibinfo
  {author} {\bibfnamefont {Q.}~\bibnamefont {Zhang}}, \ and\ \bibinfo {author}
  {\bibfnamefont {T.~J.}\ \bibnamefont {Cui}},\ }\bibfield  {title} {\enquote
  {\bibinfo {title} {{Ultra Wideband Polarization-Selective Conversions of
  Electromagnetic Waves by Metasurface under Large-Range Incident Angles.}}}\
  }\href {\doibase 10.1038/srep12476} {\bibfield  {journal} {\bibinfo
  {journal} {Sci. Rep.}\ }\textbf {\bibinfo {volume} {5}},\ \bibinfo {pages}
  {12476} (\bibinfo {year} {2015})}\BibitemShut {NoStop}%
\bibitem [{\citenamefont {Wakatsuchi}\ \emph {et~al.}(2013)\citenamefont
  {Wakatsuchi}, \citenamefont {Kim}, \citenamefont {Rushton},\ and\
  \citenamefont {Sievenpiper}}]{Wakatsuchi2013}%
  \BibitemOpen
  \bibfield  {author} {\bibinfo {author} {\bibfnamefont {H.}~\bibnamefont
  {Wakatsuchi}}, \bibinfo {author} {\bibfnamefont {S.}~\bibnamefont {Kim}},
  \bibinfo {author} {\bibfnamefont {J.~J.}\ \bibnamefont {Rushton}}, \ and\
  \bibinfo {author} {\bibfnamefont {D.~F.}\ \bibnamefont {Sievenpiper}},\
  }\bibfield  {title} {\enquote {\bibinfo {title} {Waveform-dependent absorbing
  metasurfaces},}\ }\href {\doibase 10.1103/PhysRevLett.111.245501} {\bibfield
  {journal} {\bibinfo  {journal} {Phys. Rev. Lett.}\ }\textbf {\bibinfo
  {volume} {111}},\ \bibinfo {pages} {245501} (\bibinfo {year}
  {2013})}\BibitemShut {NoStop}%
\bibitem [{\citenamefont {Asadchy}\ \emph
  {et~al.}(2015{\natexlab{b}})\citenamefont {Asadchy}, \citenamefont
  {Faniayeu}, \citenamefont {Ra'di}, \citenamefont {Khakhomov}, \citenamefont
  {Semchenko},\ and\ \citenamefont {Tretyakov}}]{Asadchy2015_1}%
  \BibitemOpen
  \bibfield  {author} {\bibinfo {author} {\bibfnamefont {V.~S.}\ \bibnamefont
  {Asadchy}}, \bibinfo {author} {\bibfnamefont {I.~A.}\ \bibnamefont
  {Faniayeu}}, \bibinfo {author} {\bibfnamefont {Y.}~\bibnamefont {Ra'di}},
  \bibinfo {author} {\bibfnamefont {S.~A.}\ \bibnamefont {Khakhomov}}, \bibinfo
  {author} {\bibfnamefont {I.~V.}\ \bibnamefont {Semchenko}}, \ and\ \bibinfo
  {author} {\bibfnamefont {S.~A.}\ \bibnamefont {Tretyakov}},\ }\bibfield
  {title} {\enquote {\bibinfo {title} {Broadband reflectionless metasheets:
  Frequency-selective transmission and perfect absorption},}\ }\href {\doibase
  10.1103/PhysRevX.5.031005} {\bibfield  {journal} {\bibinfo  {journal} {Phys.
  Rev. X}\ }\textbf {\bibinfo {volume} {5}},\ \bibinfo {pages} {031005}
  (\bibinfo {year} {2015}{\natexlab{b}})}\BibitemShut {NoStop}%
\bibitem [{\citenamefont {Ra'di}\ \emph {et~al.}(2015)\citenamefont {Ra'di},
  \citenamefont {Simovski},\ and\ \citenamefont {Tretyakov}}]{Radi2015}%
  \BibitemOpen
  \bibfield  {author} {\bibinfo {author} {\bibfnamefont {Y.}~\bibnamefont
  {Ra'di}}, \bibinfo {author} {\bibfnamefont {C.~R.}\ \bibnamefont {Simovski}},
  \ and\ \bibinfo {author} {\bibfnamefont {S.~A.}\ \bibnamefont {Tretyakov}},\
  }\bibfield  {title} {\enquote {\bibinfo {title} {Thin perfect absorbers for
  electromagnetic waves: Theory, design, and realizations},}\ }\href {\doibase
  10.1103/PhysRevApplied.3.037001} {\bibfield  {journal} {\bibinfo  {journal}
  {Phys. Rev. Appl.}\ }\textbf {\bibinfo {volume} {3}},\ \bibinfo {pages}
  {037001} (\bibinfo {year} {2015})}\BibitemShut {NoStop}%
\bibitem [{\citenamefont {Monti}\ \emph {et~al.}(2012)\citenamefont {Monti},
  \citenamefont {Soric}, \citenamefont {Alu}, \citenamefont {Bilotti},
  \citenamefont {Toscano},\ and\ \citenamefont {Vegni}}]{Monti2012}%
  \BibitemOpen
  \bibfield  {author} {\bibinfo {author} {\bibfnamefont {A.}~\bibnamefont
  {Monti}}, \bibinfo {author} {\bibfnamefont {J.}~\bibnamefont {Soric}},
  \bibinfo {author} {\bibfnamefont {A.}~\bibnamefont {Alu}}, \bibinfo {author}
  {\bibfnamefont {F.}~\bibnamefont {Bilotti}}, \bibinfo {author} {\bibfnamefont
  {A.}~\bibnamefont {Toscano}}, \ and\ \bibinfo {author} {\bibfnamefont
  {L.}~\bibnamefont {Vegni}},\ }\bibfield  {title} {\enquote {\bibinfo {title}
  {{Overcoming mutual blockage between neighboring dipole antennas using a
  low-profile patterned metasurface}},}\ }\href {\doibase
  10.1109/LAWP.2012.2229102} {\bibfield  {journal} {\bibinfo  {journal} {IEEE
  Antennas Wireless Propag. Lett.}\ }\textbf {\bibinfo {volume} {11}},\
  \bibinfo {pages} {1414--1417} (\bibinfo {year} {2012})}\BibitemShut {NoStop}%
\bibitem [{\citenamefont {Sounas}\ \emph {et~al.}(2015)\citenamefont {Sounas},
  \citenamefont {Fleury},\ and\ \citenamefont {Al{\`{u}}}}]{Sounas2015}%
  \BibitemOpen
  \bibfield  {author} {\bibinfo {author} {\bibfnamefont {D.~L.}\ \bibnamefont
  {Sounas}}, \bibinfo {author} {\bibfnamefont {R.}~\bibnamefont {Fleury}}, \
  and\ \bibinfo {author} {\bibfnamefont {A.}~\bibnamefont {Al{\`{u}}}},\
  }\bibfield  {title} {\enquote {\bibinfo {title} {{Unidirectional Cloaking
  Based on Metasurfaces with Balanced Loss and Gain}},}\ }\href {\doibase
  10.1103/PhysRevApplied.4.014005} {\bibfield  {journal} {\bibinfo  {journal}
  {Phys. Rev. Appl.}\ }\textbf {\bibinfo {volume} {4}},\ \bibinfo {pages}
  {014005} (\bibinfo {year} {2015})}\BibitemShut {NoStop}%
\bibitem [{\citenamefont {Vellucci}\ \emph {et~al.}(2017)\citenamefont
  {Vellucci}, \citenamefont {Monti}, \citenamefont {Toscano},\ and\
  \citenamefont {Bilotti}}]{Vellucci2017}%
  \BibitemOpen
  \bibfield  {author} {\bibinfo {author} {\bibfnamefont {S.}~\bibnamefont
  {Vellucci}}, \bibinfo {author} {\bibfnamefont {A.}~\bibnamefont {Monti}},
  \bibinfo {author} {\bibfnamefont {A.}~\bibnamefont {Toscano}}, \ and\
  \bibinfo {author} {\bibfnamefont {F.}~\bibnamefont {Bilotti}},\ }\bibfield
  {title} {\enquote {\bibinfo {title} {{Scattering Manipulation and Camouflage
  of Electrically Small Objects through Metasurfaces}},}\ }\href {\doibase
  10.1103/PhysRevApplied.7.034032} {\bibfield  {journal} {\bibinfo  {journal}
  {Phys. Rev. Appl.}\ }\textbf {\bibinfo {volume} {7}},\ \bibinfo {pages}
  {034032} (\bibinfo {year} {2017})}\BibitemShut {NoStop}%
\bibitem [{\citenamefont {Pfeiffer}\ and\ \citenamefont
  {Grbic}(2015)}]{Pfeiffer2015_1}%
  \BibitemOpen
  \bibfield  {author} {\bibinfo {author} {\bibfnamefont {C.}~\bibnamefont
  {Pfeiffer}}\ and\ \bibinfo {author} {\bibfnamefont {A.}~\bibnamefont
  {Grbic}},\ }\bibfield  {title} {\enquote {\bibinfo {title} {Planar lens
  antennas of subwavelength thickness: Collimating leaky-waves with
  metasurfaces},}\ }\href {\doibase 10.1109/TAP.2015.2422832} {\bibfield
  {journal} {\bibinfo  {journal} {IEEE Trans. Antennas Propag.}\ }\textbf
  {\bibinfo {volume} {63}},\ \bibinfo {pages} {3248--3253} (\bibinfo {year}
  {2015})}\BibitemShut {NoStop}%
\bibitem [{\citenamefont {Epstein}\ \emph {et~al.}(2016)\citenamefont
  {Epstein}, \citenamefont {Wong},\ and\ \citenamefont
  {Eleftheriades}}]{Epstein2016}%
  \BibitemOpen
  \bibfield  {author} {\bibinfo {author} {\bibfnamefont {A.}~\bibnamefont
  {Epstein}}, \bibinfo {author} {\bibfnamefont {J.~P.~S.}\ \bibnamefont
  {Wong}}, \ and\ \bibinfo {author} {\bibfnamefont {G.~V.}\ \bibnamefont
  {Eleftheriades}},\ }\bibfield  {title} {\enquote {\bibinfo {title}
  {Cavity-excited {H}uygens' metasurface antennas for near-unity aperture
  efficiency from arbitrarily large apertures},}\ }\href@noop {} {\bibfield
  {journal} {\bibinfo  {journal} {Nat. Commun.}\ }\textbf {\bibinfo {volume}
  {7}},\ \bibinfo {pages} {10360} (\bibinfo {year} {2016})}\BibitemShut
  {NoStop}%
\bibitem [{\citenamefont {Epstein}\ and\ \citenamefont
  {Eleftheriades}(2017)}]{Epstein2017}%
  \BibitemOpen
  \bibfield  {author} {\bibinfo {author} {\bibfnamefont {A.}~\bibnamefont
  {Epstein}}\ and\ \bibinfo {author} {\bibfnamefont {G.~V.}\ \bibnamefont
  {Eleftheriades}},\ }\bibfield  {title} {\enquote {\bibinfo {title} {Arbitrary
  antenna arrays without feed networks based on cavity-excited
  omega-bianisotropic metasurfaces},}\ }\href {\doibase
  10.1109/TAP.2017.2670358} {\bibfield  {journal} {\bibinfo  {journal} {IEEE
  Trans. Antennas Propag.}\ }\textbf {\bibinfo {volume} {65}},\ \bibinfo
  {pages} {1749--1756} (\bibinfo {year} {2017})}\BibitemShut {NoStop}%
\bibitem [{\citenamefont {Raeker}\ and\ \citenamefont
  {Rudolph}(2016)}]{Raeker2016}%
  \BibitemOpen
  \bibfield  {author} {\bibinfo {author} {\bibfnamefont {B.~O.}\ \bibnamefont
  {Raeker}}\ and\ \bibinfo {author} {\bibfnamefont {S.~M.}\ \bibnamefont
  {Rudolph}},\ }\bibfield  {title} {\enquote {\bibinfo {title} {Arbitrary
  transformation of radiation patterns using a spherical impedance
  metasurface},}\ }\href {\doibase 10.1109/TAP.2016.2618481} {\bibfield
  {journal} {\bibinfo  {journal} {IEEE Trans. Antennas Propag.}\ }\textbf
  {\bibinfo {volume} {64}},\ \bibinfo {pages} {5243--5250} (\bibinfo {year}
  {2016})}\BibitemShut {NoStop}%
\bibitem [{\citenamefont {Raeker}\ and\ \citenamefont
  {Rudolph}(2017)}]{Raeker2017}%
  \BibitemOpen
  \bibfield  {author} {\bibinfo {author} {\bibfnamefont {B.~O.}\ \bibnamefont
  {Raeker}}\ and\ \bibinfo {author} {\bibfnamefont {S.~M.}\ \bibnamefont
  {Rudolph}},\ }\bibfield  {title} {\enquote {\bibinfo {title} {Verification of
  arbitrary radiation pattern control using a cylindrical impedance
  metasurface},}\ }\href {\doibase 10.1109/LAWP.2016.2616106} {\bibfield
  {journal} {\bibinfo  {journal} {IEEE Antennas Wireless Propag. Lett.}\
  }\textbf {\bibinfo {volume} {16}},\ \bibinfo {pages} {995--998} (\bibinfo
  {year} {2017})}\BibitemShut {NoStop}%
\bibitem [{\citenamefont {Minatti}\ \emph {et~al.}(2016)\citenamefont
  {Minatti}, \citenamefont {Caminita}, \citenamefont {Martini}, \citenamefont
  {Sabbadini},\ and\ \citenamefont {Maci}}]{Minatti2016_1}%
  \BibitemOpen
  \bibfield  {author} {\bibinfo {author} {\bibfnamefont {G.}~\bibnamefont
  {Minatti}}, \bibinfo {author} {\bibfnamefont {F.}~\bibnamefont {Caminita}},
  \bibinfo {author} {\bibfnamefont {E.}~\bibnamefont {Martini}}, \bibinfo
  {author} {\bibfnamefont {M.}~\bibnamefont {Sabbadini}}, \ and\ \bibinfo
  {author} {\bibfnamefont {S.}~\bibnamefont {Maci}},\ }\bibfield  {title}
  {\enquote {\bibinfo {title} {Synthesis of modulated-metasurface antennas with
  amplitude, phase and polarization control},}\ }\href {\doibase
  10.1109/TAP.2016.2589969} {\bibfield  {journal} {\bibinfo  {journal} {IEEE
  Trans. Antennas Propag.}\ }\textbf {\bibinfo {volume} {64}},\ \bibinfo
  {pages} {3907 -- 3919} (\bibinfo {year} {2016})}\BibitemShut {NoStop}%
\bibitem [{\citenamefont {Kuester}\ \emph {et~al.}(2003)\citenamefont
  {Kuester}, \citenamefont {Mohamed}, \citenamefont {Piket-May},\ and\
  \citenamefont {Holloway}}]{Kuester2003}%
  \BibitemOpen
  \bibfield  {author} {\bibinfo {author} {\bibfnamefont {E.F.}\ \bibnamefont
  {Kuester}}, \bibinfo {author} {\bibfnamefont {M.A.}\ \bibnamefont {Mohamed}},
  \bibinfo {author} {\bibfnamefont {M.}~\bibnamefont {Piket-May}}, \ and\
  \bibinfo {author} {\bibfnamefont {C.L.}\ \bibnamefont {Holloway}},\
  }\bibfield  {title} {\enquote {\bibinfo {title} {{Averaged transition
  conditions for electromagnetic fields at a metafilm}},}\ }\href {\doibase
  10.1109/TAP.2003.817560} {\bibfield  {journal} {\bibinfo  {journal} {IEEE
  Trans. Antennas Propag.}\ }\textbf {\bibinfo {volume} {51}},\ \bibinfo
  {pages} {2641--2651} (\bibinfo {year} {2003})}\BibitemShut {NoStop}%
\bibitem [{\citenamefont {Tretyakov}(2003)}]{Tretyakov2003}%
  \BibitemOpen
  \bibfield  {author} {\bibinfo {author} {\bibfnamefont {Sergei}\ \bibnamefont
  {Tretyakov}},\ }\href@noop {} {\emph {\bibinfo {title} {Analytical Modeling
  in Applied Electromagnetics}}}\ (\bibinfo  {publisher} {Artech House},\
  \bibinfo {year} {2003})\BibitemShut {NoStop}%
\bibitem [{\citenamefont {Epstein}\ and\ \citenamefont
  {Eleftheriades}(2016{\natexlab{c}})}]{Epstein2016_2}%
  \BibitemOpen
  \bibfield  {author} {\bibinfo {author} {\bibfnamefont {A.}~\bibnamefont
  {Epstein}}\ and\ \bibinfo {author} {\bibfnamefont {G.~V.}\ \bibnamefont
  {Eleftheriades}},\ }\bibfield  {title} {\enquote {\bibinfo {title} {Huygens'
  metasurfaces via the equivalence principle: design and applications},}\
  }\href@noop {} {\bibfield  {journal} {\bibinfo  {journal} {J. Opt. Soc. Am.
  B}\ }\textbf {\bibinfo {volume} {33}},\ \bibinfo {pages} {A31--A50} (\bibinfo
  {year} {2016}{\natexlab{c}})}\BibitemShut {NoStop}%
\bibitem [{\citenamefont {Estakhri}\ and\ \citenamefont
  {Al\`{u}}(2016)}]{Estakhri2016}%
  \BibitemOpen
  \bibfield  {author} {\bibinfo {author} {\bibfnamefont {N.~Mohammadi}\
  \bibnamefont {Estakhri}}\ and\ \bibinfo {author} {\bibfnamefont
  {A.}~\bibnamefont {Al\`{u}}},\ }\bibfield  {title} {\enquote {\bibinfo
  {title} {Recent progress in gradient metasurfaces},}\ }\href {\doibase
  10.1364/JOSAB.33.000A21} {\bibfield  {journal} {\bibinfo  {journal} {J. Opt.
  Soc. Am. B}\ }\textbf {\bibinfo {volume} {33}},\ \bibinfo {pages} {A21--A30}
  (\bibinfo {year} {2016})}\BibitemShut {NoStop}%
\bibitem [{\citenamefont {Epstein}\ and\ \citenamefont
  {Eleftheriades}(2014{\natexlab{a}})}]{Epstein2014}%
  \BibitemOpen
  \bibfield  {author} {\bibinfo {author} {\bibfnamefont {A.}~\bibnamefont
  {Epstein}}\ and\ \bibinfo {author} {\bibfnamefont {G.~V.}\ \bibnamefont
  {Eleftheriades}},\ }\bibfield  {title} {\enquote {\bibinfo {title} {Passive
  lossless {Huygens} metasurfaces for conversion of arbitrary source field to
  directive radiation},}\ }\href {\doibase 10.1109/TAP.2014.2354419} {\bibfield
   {journal} {\bibinfo  {journal} {IEEE Trans. Antennas Propag.}\ }\textbf
  {\bibinfo {volume} {62}},\ \bibinfo {pages} {5680--5695} (\bibinfo {year}
  {2014}{\natexlab{a}})}\BibitemShut {NoStop}%
\bibitem [{\citenamefont {Ranjbar}\ and\ \citenamefont
  {Grbic}(2017)}]{Ranjbar2017}%
  \BibitemOpen
  \bibfield  {author} {\bibinfo {author} {\bibfnamefont {A.}~\bibnamefont
  {Ranjbar}}\ and\ \bibinfo {author} {\bibfnamefont {A.}~\bibnamefont
  {Grbic}},\ }\bibfield  {title} {\enquote {\bibinfo {title} {Analysis and
  synthesis of cascaded metasurfaces using wave matrices},}\ }\href {\doibase
  10.1103/PhysRevB.95.205114} {\bibfield  {journal} {\bibinfo  {journal} {Phys.
  Rev. B}\ }\textbf {\bibinfo {volume} {95}},\ \bibinfo {pages} {205114}
  (\bibinfo {year} {2017})}\BibitemShut {NoStop}%
\bibitem [{\citenamefont {Alaee}\ \emph
  {et~al.}(2015{\natexlab{a}})\citenamefont {Alaee}, \citenamefont {Albooyeh},
  \citenamefont {Yazdi}, \citenamefont {Komjani}, \citenamefont {Simovski},
  \citenamefont {Lederer},\ and\ \citenamefont {Rockstuhl}}]{Alaee2015}%
  \BibitemOpen
  \bibfield  {author} {\bibinfo {author} {\bibfnamefont {R.}~\bibnamefont
  {Alaee}}, \bibinfo {author} {\bibfnamefont {M.}~\bibnamefont {Albooyeh}},
  \bibinfo {author} {\bibfnamefont {M.}~\bibnamefont {Yazdi}}, \bibinfo
  {author} {\bibfnamefont {N.}~\bibnamefont {Komjani}}, \bibinfo {author}
  {\bibfnamefont {C.}~\bibnamefont {Simovski}}, \bibinfo {author}
  {\bibfnamefont {F.}~\bibnamefont {Lederer}}, \ and\ \bibinfo {author}
  {\bibfnamefont {C.}~\bibnamefont {Rockstuhl}},\ }\bibfield  {title} {\enquote
  {\bibinfo {title} {Magnetoelectric coupling in nonidentical plasmonic
  nanoparticles: Theory and applications},}\ }\href {\doibase
  10.1103/PhysRevB.91.115119} {\bibfield  {journal} {\bibinfo  {journal} {Phys.
  Rev. B}\ }\textbf {\bibinfo {volume} {91}},\ \bibinfo {pages} {115119}
  (\bibinfo {year} {2015}{\natexlab{a}})}\BibitemShut {NoStop}%
\bibitem [{\citenamefont {Alaee}\ \emph
  {et~al.}(2015{\natexlab{b}})\citenamefont {Alaee}, \citenamefont {Albooyeh},
  \citenamefont {Rahimzadegan}, \citenamefont {Mirmoosa}, \citenamefont
  {Kivshar},\ and\ \citenamefont {Rockstuhl}}]{Alaee2015_1}%
  \BibitemOpen
  \bibfield  {author} {\bibinfo {author} {\bibfnamefont {R.}~\bibnamefont
  {Alaee}}, \bibinfo {author} {\bibfnamefont {M.}~\bibnamefont {Albooyeh}},
  \bibinfo {author} {\bibfnamefont {A.}~\bibnamefont {Rahimzadegan}}, \bibinfo
  {author} {\bibfnamefont {M.~S.}\ \bibnamefont {Mirmoosa}}, \bibinfo {author}
  {\bibfnamefont {Y.~S.}\ \bibnamefont {Kivshar}}, \ and\ \bibinfo {author}
  {\bibfnamefont {C.}~\bibnamefont {Rockstuhl}},\ }\bibfield  {title} {\enquote
  {\bibinfo {title} {{All-dielectric reciprocal bianisotropic
  nanoparticles}},}\ }\href {\doibase 10.1103/PhysRevB.92.245130} {\bibfield
  {journal} {\bibinfo  {journal} {Phys. Rev. B}\ }\textbf {\bibinfo {volume}
  {92}},\ \bibinfo {pages} {245130} (\bibinfo {year}
  {2015}{\natexlab{b}})}\BibitemShut {NoStop}%
\bibitem [{\citenamefont {Odit}\ \emph {et~al.}(2016)\citenamefont {Odit},
  \citenamefont {Kapitanova}, \citenamefont {Belov}, \citenamefont {Alaee},
  \citenamefont {Rockstuhl},\ and\ \citenamefont {Kivshar}}]{Odit2016}%
  \BibitemOpen
  \bibfield  {author} {\bibinfo {author} {\bibfnamefont {M.}~\bibnamefont
  {Odit}}, \bibinfo {author} {\bibfnamefont {P.}~\bibnamefont {Kapitanova}},
  \bibinfo {author} {\bibfnamefont {P.}~\bibnamefont {Belov}}, \bibinfo
  {author} {\bibfnamefont {R.}~\bibnamefont {Alaee}}, \bibinfo {author}
  {\bibfnamefont {C.}~\bibnamefont {Rockstuhl}}, \ and\ \bibinfo {author}
  {\bibfnamefont {Y.~S.}\ \bibnamefont {Kivshar}},\ }\bibfield  {title}
  {\enquote {\bibinfo {title} {{Experimental realisation of all-dielectric
  bianisotropic metasurfaces}},}\ }\href {\doibase 10.1063/1.4953023}
  {\bibfield  {journal} {\bibinfo  {journal} {Appl. Phys. Lett.}\ }\textbf
  {\bibinfo {volume} {108}},\ \bibinfo {pages} {221903} (\bibinfo {year}
  {2016})}\BibitemShut {NoStop}%
\bibitem [{\citenamefont {Kim}\ and\ \citenamefont
  {Eleftheriades}(2016)}]{Kim2016}%
  \BibitemOpen
  \bibfield  {author} {\bibinfo {author} {\bibfnamefont {M.}~\bibnamefont
  {Kim}}\ and\ \bibinfo {author} {\bibfnamefont {G.~V.}\ \bibnamefont
  {Eleftheriades}},\ }\bibfield  {title} {\enquote {\bibinfo {title} {{Highly
  efficient all-dielectric optical tensor impedance metasurfaces for chiral
  polarization control}},}\ }\href {\doibase 10.1364/OL.41.004831} {\bibfield
  {journal} {\bibinfo  {journal} {Opt. Lett.}\ }\textbf {\bibinfo {volume}
  {41}},\ \bibinfo {pages} {4831} (\bibinfo {year} {2016})}\BibitemShut
  {NoStop}%
\bibitem [{\citenamefont {Asadchy}\ \emph
  {et~al.}(2016{\natexlab{b}})\citenamefont {Asadchy}, \citenamefont
  {Albooyeh},\ and\ \citenamefont {Tretyakov}}]{Asadchy2016_2}%
  \BibitemOpen
  \bibfield  {author} {\bibinfo {author} {\bibfnamefont {V.}~\bibnamefont
  {Asadchy}}, \bibinfo {author} {\bibfnamefont {M.}~\bibnamefont {Albooyeh}}, \
  and\ \bibinfo {author} {\bibfnamefont {S.}~\bibnamefont {Tretyakov}},\
  }\bibfield  {title} {\enquote {\bibinfo {title} {{Optical metamirror:
  all-dielectric frequency-selective mirror with fully controllable reflection
  phase}},}\ }\href {\doibase 10.1364/JOSAB.33.000A16} {\bibfield  {journal}
  {\bibinfo  {journal} {J. Opt. Soc. Am. B}\ }\textbf {\bibinfo {volume}
  {33}},\ \bibinfo {pages} {A16} (\bibinfo {year}
  {2016}{\natexlab{b}})}\BibitemShut {NoStop}%
\bibitem [{\citenamefont {Sounas}\ \emph {et~al.}(2016)\citenamefont {Sounas},
  \citenamefont {Estakhri},\ and\ \citenamefont {Alu}}]{Sounas2016}%
  \BibitemOpen
  \bibfield  {author} {\bibinfo {author} {\bibfnamefont {D.~L.}\ \bibnamefont
  {Sounas}}, \bibinfo {author} {\bibfnamefont {N.~Mohammadi}\ \bibnamefont
  {Estakhri}}, \ and\ \bibinfo {author} {\bibfnamefont {A.}~\bibnamefont
  {Alu}},\ }\bibfield  {title} {\enquote {\bibinfo {title} {{Metasurfaces with
  engineered reflection and transmission: Optimal designs through coupled-mode
  analysis}},}\ }in\ \href {\doibase 10.1109/MetaMaterials.2016.7746394} {\emph
  {\bibinfo {booktitle} {2016 10th International Congress on Advanced
  Electromagnetic Materials in Microwaves and Optics (METAMATERIALS)}}}\
  (\bibinfo {year} {2016})\ pp.\ \bibinfo {pages} {346--348}\BibitemShut
  {NoStop}%
\bibitem [{\citenamefont {Wong}\ \emph {et~al.}(2017)\citenamefont {Wong},
  \citenamefont {Christian},\ and\ \citenamefont {Eleftheriades}}]{Wong2017}%
  \BibitemOpen
  \bibfield  {author} {\bibinfo {author} {\bibfnamefont {A.~M.~H.}\
  \bibnamefont {Wong}}, \bibinfo {author} {\bibfnamefont {P.}~\bibnamefont
  {Christian}}, \ and\ \bibinfo {author} {\bibfnamefont {G.~V.}\ \bibnamefont
  {Eleftheriades}},\ }\bibfield  {title} {\enquote {\bibinfo {title} {{Binary
  Huygens' metasurface: A simple and efficient retroreflector at near-grazing
  angles}},}\ }in\ \href {\doibase 10.1109/USNC-URSI-NRSM.2017.7878269} {\emph
  {\bibinfo {booktitle} {2017 United States National Committee of URSI National
  Radio Science Meeting (USNC-URSI NRSM)}}}\ (\bibinfo {year}
  {2017})\BibitemShut {NoStop}%
\bibitem [{\citenamefont {Memarian}\ \emph {et~al.}(2017)\citenamefont
  {Memarian}, \citenamefont {Li}, \citenamefont {Morimoto},\ and\ \citenamefont
  {Itoh}}]{Memarian2017}%
  \BibitemOpen
  \bibfield  {author} {\bibinfo {author} {\bibfnamefont {M.}~\bibnamefont
  {Memarian}}, \bibinfo {author} {\bibfnamefont {X.}~\bibnamefont {Li}},
  \bibinfo {author} {\bibfnamefont {Y.}~\bibnamefont {Morimoto}}, \ and\
  \bibinfo {author} {\bibfnamefont {T.}~\bibnamefont {Itoh}},\ }\bibfield
  {title} {\enquote {\bibinfo {title} {{Wide-band/angle Blazed Surfaces using
  Multiple Coupled Blazing Resonances}},}\ }\href {\doibase 10.1038/srep42286}
  {\bibfield  {journal} {\bibinfo  {journal} {Sci. Rep.}\ }\textbf {\bibinfo
  {volume} {7}},\ \bibinfo {pages} {42286} (\bibinfo {year}
  {2017})}\BibitemShut {NoStop}%
\bibitem [{\citenamefont {Paniagua-Dominguez}\ \emph
  {et~al.}(2017)\citenamefont {Paniagua-Dominguez}, \citenamefont {Yu},
  \citenamefont {Khaidarov}, \citenamefont {Bakker}, \citenamefont {Liang},
  \citenamefont {Fu},\ and\ \citenamefont {Kuznetsov}}]{PaniaguaDominguez2017}%
  \BibitemOpen
  \bibfield  {author} {\bibinfo {author} {\bibfnamefont {R.}~\bibnamefont
  {Paniagua-Dominguez}}, \bibinfo {author} {\bibfnamefont {Y.~F.}\ \bibnamefont
  {Yu}}, \bibinfo {author} {\bibfnamefont {E.}~\bibnamefont {Khaidarov}},
  \bibinfo {author} {\bibfnamefont {R.~M.}\ \bibnamefont {Bakker}}, \bibinfo
  {author} {\bibfnamefont {X.}~\bibnamefont {Liang}}, \bibinfo {author}
  {\bibfnamefont {Y.~H.}\ \bibnamefont {Fu}}, \ and\ \bibinfo {author}
  {\bibfnamefont {A.~I.}\ \bibnamefont {Kuznetsov}},\ }\bibfield  {title}
  {\enquote {\bibinfo {title} {{A Metalens with Near-Unity Numerical
  Aperture}},}\ }\href {http://arxiv.org/abs/1705.00895} {\  (\bibinfo {year}
  {2017})},\ \Eprint {http://arxiv.org/abs/1705.00895} {arXiv:1705.00895}
  \BibitemShut {NoStop}%
\bibitem [{\citenamefont {Ra'di}\ \emph {et~al.}(2017)\citenamefont {Ra'di},
  \citenamefont {Sounas},\ and\ \citenamefont {Al{\`{u}}}}]{Radi2017}%
  \BibitemOpen
  \bibfield  {author} {\bibinfo {author} {\bibfnamefont {Y.}~\bibnamefont
  {Ra'di}}, \bibinfo {author} {\bibfnamefont {D.~L.}\ \bibnamefont {Sounas}}, \
  and\ \bibinfo {author} {\bibfnamefont {A.}~\bibnamefont {Al{\`{u}}}},\
  }\bibfield  {title} {\enquote {\bibinfo {title} {{Metagratings: Beyond the
  Limits of Graded Metasurfaces for Wave Front Control}},}\ }\href {\doibase
  10.1103/PhysRevLett.119.067404} {\bibfield  {journal} {\bibinfo  {journal}
  {Phys. Rev. Lett.}\ }\textbf {\bibinfo {volume} {119}},\ \bibinfo {pages}
  {067404} (\bibinfo {year} {2017})}\BibitemShut {NoStop}%
\bibitem [{\citenamefont {Wong}\ and\ \citenamefont
  {Eleftheriades}(2017)}]{Wong2017_1}%
  \BibitemOpen
  \bibfield  {author} {\bibinfo {author} {\bibfnamefont {A.~M.~H.}\
  \bibnamefont {Wong}}\ and\ \bibinfo {author} {\bibfnamefont {G.~V.}\
  \bibnamefont {Eleftheriades}},\ }\bibfield  {title} {\enquote {\bibinfo
  {title} {{Perfect Anomalous Reflection with an Aggressively Discretized
  Huygens' Metasurface}},}\ }\href {http://arxiv.org/abs/1706.02765} {\
  (\bibinfo {year} {2017})},\ \Eprint {http://arxiv.org/abs/1706.02765}
  {arXiv:1706.02765} \BibitemShut {NoStop}%
\bibitem [{\citenamefont {D{\'{i}}az-Rubio}\ \emph {et~al.}(2017)\citenamefont
  {D{\'{i}}az-Rubio}, \citenamefont {Asadchy}, \citenamefont {Elsakka},\ and\
  \citenamefont {Tretyakov}}]{DiazRubio2017}%
  \BibitemOpen
  \bibfield  {author} {\bibinfo {author} {\bibfnamefont {A.}~\bibnamefont
  {D{\'{i}}az-Rubio}}, \bibinfo {author} {\bibfnamefont {V.~S.}\ \bibnamefont
  {Asadchy}}, \bibinfo {author} {\bibfnamefont {A.}~\bibnamefont {Elsakka}}, \
  and\ \bibinfo {author} {\bibfnamefont {S.~A.}\ \bibnamefont {Tretyakov}},\
  }\bibfield  {title} {\enquote {\bibinfo {title} {{From the generalized
  reflection law to the realization of perfect anomalous reflectors}},}\ }\href
  {\doibase 10.1126/sciadv.1602714} {\bibfield  {journal} {\bibinfo  {journal}
  {Sci. Adv.}\ }\textbf {\bibinfo {volume} {3}},\ \bibinfo {pages} {e1602714}
  (\bibinfo {year} {2017})}\BibitemShut {NoStop}%
\bibitem [{\citenamefont {Asadchy}\ \emph {et~al.}(2017)\citenamefont
  {Asadchy}, \citenamefont {Wickberg}, \citenamefont {D{\'{i}}az-Rubio},\ and\
  \citenamefont {Wegener}}]{Asadchy2017}%
  \BibitemOpen
  \bibfield  {author} {\bibinfo {author} {\bibfnamefont {V.~S.}\ \bibnamefont
  {Asadchy}}, \bibinfo {author} {\bibfnamefont {A.}~\bibnamefont {Wickberg}},
  \bibinfo {author} {\bibfnamefont {A.}~\bibnamefont {D{\'{i}}az-Rubio}}, \
  and\ \bibinfo {author} {\bibfnamefont {M.}~\bibnamefont {Wegener}},\
  }\bibfield  {title} {\enquote {\bibinfo {title} {{Eliminating Scattering Loss
  in Anomalously Reflecting Optical Metasurfaces}},}\ }\href {\doibase
  10.1021/acsphotonics.7b00213} {\bibfield  {journal} {\bibinfo  {journal} {ACS
  Photonics}\ }\textbf {\bibinfo {volume} {4}},\ \bibinfo {pages} {1264--1270}
  (\bibinfo {year} {2017})}\BibitemShut {NoStop}%
\bibitem [{\citenamefont {Epstein}\ and\ \citenamefont
  {Eleftheriades}(2014{\natexlab{b}})}]{Epstein2014_2}%
  \BibitemOpen
  \bibfield  {author} {\bibinfo {author} {\bibfnamefont {A.}~\bibnamefont
  {Epstein}}\ and\ \bibinfo {author} {\bibfnamefont {G.~V.}\ \bibnamefont
  {Eleftheriades}},\ }\bibfield  {title} {\enquote {\bibinfo {title}
  {{F}loquet-{B}loch analysis of refracting {H}uygens metasurfaces},}\
  }\href@noop {} {\bibfield  {journal} {\bibinfo  {journal} {Phys. Rev. B}\
  }\textbf {\bibinfo {volume} {90}},\ \bibinfo {pages} {235127} (\bibinfo
  {year} {2014}{\natexlab{b}})}\BibitemShut {NoStop}%
\bibitem [{\citenamefont {Perry}\ \emph {et~al.}(1995)\citenamefont {Perry},
  \citenamefont {Shannon}, \citenamefont {Shults}, \citenamefont {Boyd},
  \citenamefont {Britten}, \citenamefont {Decker},\ and\ \citenamefont
  {Shore}}]{Perry1995}%
  \BibitemOpen
  \bibfield  {author} {\bibinfo {author} {\bibfnamefont {M.~D.}\ \bibnamefont
  {Perry}}, \bibinfo {author} {\bibfnamefont {C.}~\bibnamefont {Shannon}},
  \bibinfo {author} {\bibfnamefont {E.}~\bibnamefont {Shults}}, \bibinfo
  {author} {\bibfnamefont {R.~D.}\ \bibnamefont {Boyd}}, \bibinfo {author}
  {\bibfnamefont {J.~A.}\ \bibnamefont {Britten}}, \bibinfo {author}
  {\bibfnamefont {D.}~\bibnamefont {Decker}}, \ and\ \bibinfo {author}
  {\bibfnamefont {B.~W.}\ \bibnamefont {Shore}},\ }\bibfield  {title} {\enquote
  {\bibinfo {title} {{High-efficiency multilayer dielectric diffraction
  gratings}},}\ }\href {\doibase 10.1364/OL.20.000940} {\bibfield  {journal}
  {\bibinfo  {journal} {Opt. Lett.}\ }\textbf {\bibinfo {volume} {20}},\
  \bibinfo {pages} {940} (\bibinfo {year} {1995})}\BibitemShut {NoStop}%
\bibitem [{\citenamefont {Destouches}\ \emph {et~al.}(2005)\citenamefont
  {Destouches}, \citenamefont {Tishchenko}, \citenamefont {Pommier},
  \citenamefont {Reynaud}, \citenamefont {Parriaux}, \citenamefont {Tonchev},\
  and\ \citenamefont {Ahmed}}]{Destouches2005}%
  \BibitemOpen
  \bibfield  {author} {\bibinfo {author} {\bibfnamefont {N.}~\bibnamefont
  {Destouches}}, \bibinfo {author} {\bibfnamefont {A.~V.}\ \bibnamefont
  {Tishchenko}}, \bibinfo {author} {\bibfnamefont {J.~C.}\ \bibnamefont
  {Pommier}}, \bibinfo {author} {\bibfnamefont {S.}~\bibnamefont {Reynaud}},
  \bibinfo {author} {\bibfnamefont {O.}~\bibnamefont {Parriaux}}, \bibinfo
  {author} {\bibfnamefont {S.}~\bibnamefont {Tonchev}}, \ and\ \bibinfo
  {author} {\bibfnamefont {M.~Abdou}\ \bibnamefont {Ahmed}},\ }\bibfield
  {title} {\enquote {\bibinfo {title} {99{\%} efficiency measured in the -1st
  order of a resonant grating},}\ }\href {\doibase 10.1364/OPEX.13.003230}
  {\bibfield  {journal} {\bibinfo  {journal} {Opt. Express}\ }\textbf {\bibinfo
  {volume} {13}},\ \bibinfo {pages} {3230} (\bibinfo {year}
  {2005})}\BibitemShut {NoStop}%
\bibitem [{\citenamefont {Ito}\ and\ \citenamefont {Iizuka}(2013)}]{Ito2013}%
  \BibitemOpen
  \bibfield  {author} {\bibinfo {author} {\bibfnamefont {K.}~\bibnamefont
  {Ito}}\ and\ \bibinfo {author} {\bibfnamefont {H.}~\bibnamefont {Iizuka}},\
  }\bibfield  {title} {\enquote {\bibinfo {title} {{Highly efficient -1st-order
  reflection in Littrow mounted dielectric double-groove grating}},}\ }\href
  {\doibase 10.1063/1.4811466} {\bibfield  {journal} {\bibinfo  {journal} {AIP
  Adv.}\ }\textbf {\bibinfo {volume} {3}},\ \bibinfo {pages} {062119} (\bibinfo
  {year} {2013})}\BibitemShut {NoStop}%
\bibitem [{\citenamefont {Pfeiffer}\ \emph
  {et~al.}(2014{\natexlab{b}})\citenamefont {Pfeiffer}, \citenamefont {Emani},
  \citenamefont {Shaltout}, \citenamefont {Boltasseva}, \citenamefont
  {Shalaev},\ and\ \citenamefont {Grbic}}]{Pfeiffer2014}%
  \BibitemOpen
  \bibfield  {author} {\bibinfo {author} {\bibfnamefont {C.}~\bibnamefont
  {Pfeiffer}}, \bibinfo {author} {\bibfnamefont {N.~K.}\ \bibnamefont {Emani}},
  \bibinfo {author} {\bibfnamefont {A.~M.}\ \bibnamefont {Shaltout}}, \bibinfo
  {author} {\bibfnamefont {A.}~\bibnamefont {Boltasseva}}, \bibinfo {author}
  {\bibfnamefont {V.~M.}\ \bibnamefont {Shalaev}}, \ and\ \bibinfo {author}
  {\bibfnamefont {A.}~\bibnamefont {Grbic}},\ }\bibfield  {title} {\enquote
  {\bibinfo {title} {Efficient light bending with isotropic metamaterial
  {H}uygens' surfaces.}}\ }\href@noop {} {\bibfield  {journal} {\bibinfo
  {journal} {Nano Lett.}\ } (\bibinfo {year} {2014}{\natexlab{b}})}\BibitemShut
  {NoStop}%
\bibitem [{\citenamefont {Kuznetsov}\ \emph {et~al.}(2015)\citenamefont
  {Kuznetsov}, \citenamefont {Astafev}, \citenamefont {Beruete},\ and\
  \citenamefont {Navarro-C{\'{i}}a}}]{Kuznetsov2015}%
  \BibitemOpen
  \bibfield  {author} {\bibinfo {author} {\bibfnamefont {S.~A.}\ \bibnamefont
  {Kuznetsov}}, \bibinfo {author} {\bibfnamefont {M.~A.}\ \bibnamefont
  {Astafev}}, \bibinfo {author} {\bibfnamefont {M.}~\bibnamefont {Beruete}}, \
  and\ \bibinfo {author} {\bibfnamefont {M.}~\bibnamefont
  {Navarro-C{\'{i}}a}},\ }\bibfield  {title} {\enquote {\bibinfo {title}
  {{Planar Holographic Metasurfaces for Terahertz Focusing}},}\ }\href
  {\doibase 10.1038/srep07738} {\bibfield  {journal} {\bibinfo  {journal} {Sci.
  Rep.}\ }\textbf {\bibinfo {volume} {5}},\ \bibinfo {pages} {7738} (\bibinfo
  {year} {2015})}\BibitemShut {NoStop}%
\bibitem [{\citenamefont {Chang}\ \emph {et~al.}(2017)\citenamefont {Chang},
  \citenamefont {Headland}, \citenamefont {Abbott}, \citenamefont
  {Withayachumnankul},\ and\ \citenamefont {Chen}}]{Chang2017}%
  \BibitemOpen
  \bibfield  {author} {\bibinfo {author} {\bibfnamefont {Ch.-Ch.}\ \bibnamefont
  {Chang}}, \bibinfo {author} {\bibfnamefont {D.}~\bibnamefont {Headland}},
  \bibinfo {author} {\bibfnamefont {D.}~\bibnamefont {Abbott}}, \bibinfo
  {author} {\bibfnamefont {W.}~\bibnamefont {Withayachumnankul}}, \ and\
  \bibinfo {author} {\bibfnamefont {H.-T.}\ \bibnamefont {Chen}},\ }\bibfield
  {title} {\enquote {\bibinfo {title} {{Demonstration of a highly efficient
  terahertz flat lens employing tri-layer metasurfaces}},}\ }\href {\doibase
  10.1364/OL.42.001867} {\bibfield  {journal} {\bibinfo  {journal} {Opt.
  Lett.}\ }\textbf {\bibinfo {volume} {42}},\ \bibinfo {pages} {1867} (\bibinfo
  {year} {2017})}\BibitemShut {NoStop}%
\bibitem [{\citenamefont {Wait}(1954)}]{Wait1954}%
  \BibitemOpen
  \bibfield  {author} {\bibinfo {author} {\bibfnamefont {J.~R.}\ \bibnamefont
  {Wait}},\ }\bibfield  {title} {\enquote {\bibinfo {title} {{Reflection from a
  wire grid parallel to a conducting plane1}},}\ }\href {\doibase
  10.1139/p54-061} {\bibfield  {journal} {\bibinfo  {journal} {Can. J. Phys.}\
  }\textbf {\bibinfo {volume} {32}},\ \bibinfo {pages} {571--579} (\bibinfo
  {year} {1954})}\BibitemShut {NoStop}%
\bibitem [{\citenamefont {Liberal}\ \emph {et~al.}(2012)\citenamefont
  {Liberal}, \citenamefont {Nefedov}, \citenamefont {Ederra}, \citenamefont
  {Gonzalo},\ and\ \citenamefont {Tretyakov}}]{Liberal2012}%
  \BibitemOpen
  \bibfield  {author} {\bibinfo {author} {\bibfnamefont {I.}~\bibnamefont
  {Liberal}}, \bibinfo {author} {\bibfnamefont {I.~S.}\ \bibnamefont
  {Nefedov}}, \bibinfo {author} {\bibfnamefont {I.}~\bibnamefont {Ederra}},
  \bibinfo {author} {\bibfnamefont {R.}~\bibnamefont {Gonzalo}}, \ and\
  \bibinfo {author} {\bibfnamefont {S.~A.}\ \bibnamefont {Tretyakov}},\
  }\bibfield  {title} {\enquote {\bibinfo {title} {{Reconfigurable artificial
  surfaces based on impedance loaded wires close to a ground plane}},}\ }\href
  {\doibase 10.1109/TAP.2012.2186264} {\bibfield  {journal} {\bibinfo
  {journal} {IEEE Trans. Antennas Propag.}\ }\textbf {\bibinfo {volume} {60}},\
  \bibinfo {pages} {1921--1930} (\bibinfo {year} {2012})}\BibitemShut {NoStop}%
\bibitem [{\citenamefont {Felsen}\ and\ \citenamefont
  {Marcuvitz}(1973)}]{FelsenMarcuvitz1973}%
  \BibitemOpen
  \bibfield  {author} {\bibinfo {author} {\bibfnamefont {L.~B.}\ \bibnamefont
  {Felsen}}\ and\ \bibinfo {author} {\bibfnamefont {N.}~\bibnamefont
  {Marcuvitz}},\ }\href@noop {} {\emph {\bibinfo {title} {Radiation and
  Scattering of Waves}}},\ \bibinfo {edition} {1st}\ ed.\ (\bibinfo
  {publisher} {Prentice-Hall},\ \bibinfo {address} {Englewood Cliffs, N.J.},\
  \bibinfo {year} {1973})\BibitemShut {NoStop}%
\bibitem [{\citenamefont {Abramowitz}\ and\ \citenamefont
  {Stegun}(1970)}]{AbramowitzStegun1970}%
  \BibitemOpen
  \bibfield  {author} {\bibinfo {author} {\bibfnamefont {M.}~\bibnamefont
  {Abramowitz}}\ and\ \bibinfo {author} {\bibfnamefont {I.~A.}\ \bibnamefont
  {Stegun}},\ }\href@noop {} {\emph {\bibinfo {title} {Handbook of Mathematical
  Functions : with Formulas, Graphs, and Mathematical Tables}}}\ (\bibinfo
  {publisher} {Dover Publications},\ \bibinfo {address} {New York},\ \bibinfo
  {year} {1970})\BibitemShut {NoStop}%
\bibitem [{\citenamefont {Gradshtein}\ and\ \citenamefont
  {Ryzhik}(2015)}]{GradshteinRyzhik2015}%
  \BibitemOpen
  \bibfield  {author} {\bibinfo {author} {\bibfnamefont {I.~S.}\ \bibnamefont
  {Gradshtein}}\ and\ \bibinfo {author} {\bibfnamefont {I.~M}\ \bibnamefont
  {Ryzhik}},\ }\bibfield  {title} {\enquote {\bibinfo {title} {{Table of
  integrals, series, and products}},}\ \ }(\bibinfo  {publisher} {Academic
  Press},\ \bibinfo {year} {2015})\ \bibinfo {edition} {8th}\ ed.\BibitemShut
  {Stop}%
\bibitem [{\citenamefont {Chen}\ \emph {et~al.}(2017)\citenamefont {Chen},
  \citenamefont {Abdo-S{\'{a}}nchez}, \citenamefont {Epstein},\ and\
  \citenamefont {Eleftheriades}}]{Chen2017}%
  \BibitemOpen
  \bibfield  {author} {\bibinfo {author} {\bibfnamefont {M.}~\bibnamefont
  {Chen}}, \bibinfo {author} {\bibfnamefont {E.}~\bibnamefont
  {Abdo-S{\'{a}}nchez}}, \bibinfo {author} {\bibfnamefont {A.}~\bibnamefont
  {Epstein}}, \ and\ \bibinfo {author} {\bibfnamefont {G.~V.}\ \bibnamefont
  {Eleftheriades}},\ }\bibfield  {title} {\enquote {\bibinfo {title}
  {{Experimental Verification of Reflectionless Wide-Angle Refraction via a
  Bianisotropic Huygens' Metasurface}},}\ }\href@noop {} {\bibfield  {journal}
  {\bibinfo  {journal} {32nd International Union of Radio Science General
  Assembly and Scientific Symposium (URSI2017)}\ ,\ \bibinfo {pages}
  {arXiv:1703.06669}} (\bibinfo {year} {2017})}\BibitemShut {NoStop}%
\bibitem [{\citenamefont {Lee}(2003)}]{Lee2003}%
  \BibitemOpen
  \bibfield  {author} {\bibinfo {author} {\bibfnamefont {T.~H.}\ \bibnamefont
  {Lee}},\ }\href@noop {} {\emph {\bibinfo {title} {The design of CMOS
  radio-frequency integrated circuits}}}\ (\bibinfo  {publisher} {Cambridge
  university press},\ \bibinfo {year} {2003})\BibitemShut {NoStop}%
\bibitem [{\citenamefont {Gupta}\ \emph {et~al.}(1996)\citenamefont {Gupta},
  \citenamefont {Garg}, \citenamefont {Bahl},\ and\ \citenamefont
  {Bhartia}}]{Gupta1996}%
  \BibitemOpen
  \bibfield  {author} {\bibinfo {author} {\bibfnamefont {K.~C.}\ \bibnamefont
  {Gupta}}, \bibinfo {author} {\bibfnamefont {R.}~\bibnamefont {Garg}},
  \bibinfo {author} {\bibfnamefont {I.}~\bibnamefont {Bahl}}, \ and\ \bibinfo
  {author} {\bibfnamefont {P.}~\bibnamefont {Bhartia}},\ }\href@noop {} {\emph
  {\bibinfo {title} {Microstrip Lines and Slotlines}}}\ (\bibinfo  {publisher}
  {Artech House},\ \bibinfo {address} {Boston},\ \bibinfo {year}
  {1996})\BibitemShut {NoStop}%
\bibitem [{\citenamefont {Wong}\ \emph {et~al.}(2015)\citenamefont {Wong},
  \citenamefont {Epstein},\ and\ \citenamefont {Eleftheriades}}]{Wong2016}%
  \BibitemOpen
  \bibfield  {author} {\bibinfo {author} {\bibfnamefont {J.~P.~S.}\
  \bibnamefont {Wong}}, \bibinfo {author} {\bibfnamefont {A.}~\bibnamefont
  {Epstein}}, \ and\ \bibinfo {author} {\bibfnamefont {G.~V.}\ \bibnamefont
  {Eleftheriades}},\ }\bibfield  {title} {\enquote {\bibinfo {title}
  {Reflectionless wide-angle refracting metasurfaces},}\ }\href {\doibase
  10.1109/LAWP.2015.2505629} {\bibfield  {journal} {\bibinfo  {journal} {IEEE
  Antennas Wireless Propag. Lett.}\ }\textbf {\bibinfo {volume} {15}},\
  \bibinfo {pages} {1293--1296} (\bibinfo {year} {2015})}\BibitemShut {NoStop}%
\end{thebibliography}%

\end{document}